\documentclass[12pt]{article}
\usepackage{hyperref}
\usepackage[pdftex]{graphicx,color}
\usepackage{slashed}
\usepackage{amsmath,amssymb}
\usepackage{hhline}
\usepackage{subfigure}
\usepackage{epstopdf}
\usepackage{epsfig}
\usepackage{graphicx}
\usepackage{caption}
\usepackage{tabularx}

\newcommand{\braket}[2]{\left\langle #1 | #2 \right\rangle}

\def\beq{\begin{equation}}
\def\eeq#1{\label{#1}\end{equation}}
\def\eeqn{\end{equation}}

\def\beqa{\begin{eqnarray}}
\def\eeqa#1{\label{#1}\end{eqnarray}}
\def\eeqan{\end{eqnarray}}

\let\bar=\overbar

\def\Dslash{\not{\hbox{\kern-4pt $D$}}}
\def\dslash{\not{\hbox{\kern-2pt $\del$}}}

\def\msb{{\bar{\ssstyle M \kern -1pt S}}}

\def\Title#1{\begin{center} {\Large {\bf #1} } \end{center}}

\begin{document}

\Title{Lepton flavor violating meson decays \\
{\small \textsl{Talk presented at the APS Division of Particles and Fields Meeting \\
(DPF 2017), July 31-August 4, 2017, Fermilab. C170731}}}

\bigskip\bigskip


\begin{raggedright}  

{\it Derek E. Hazard \index{Hazard, D.}\\
Department of Physics and Astronomy\\
Wayne State University\\
Detroit, Michigan 48201, USA}
\bigskip\bigskip
\end{raggedright}

\begin{abstract}
We argue that lepton flavor violating (LFV) decays $M \to \ell_1 \overline \ell_2$ of meson states $M$ with 
different quantum numbers could be used to put constraints on the Wilson 
coefficients of effective operators describing LFV interactions at low energy scales. We note that the 
restricted kinematics of the two-body decay of quarkonium or a heavy quark meson allows us to select 
operators with particular quantum numbers, significantly reducing the reliance on the \textit{single operator 
dominance} assumption that is prevalent in constraining parameters of the effective LFV Lagrangian. We 
shall also argue that studies of radiative lepton flavor violating $M \to \gamma \ell_1 \overline \ell_2$ 
decays could provide important complementary access to those effective operators.
\end{abstract}

\section{Introduction}

Flavor-changing neutral current (FCNC) interactions serve as a powerful probe of physics beyond the Standard 
Model (BSM). Since no operators generate FCNCs in the Standard Model (SM) at tree level, new physics (NP) degrees of 
freedom can effectively compete with SM particles running in loop graphs, making their discovery possible.

A convenient way to describe contributions of NP at low energies is offered by effective Lagrangians. Lepton 
flavor violating (LFV) meson decays can be used to constrain Wilson coefficients (WCs) of effective operators
describing interactions at low energy scales.  The restricted kinematics of 2-body decays allows for the selection of operators with
certain quantum numbers, reducing the reliance on the \textit{single operator dominance} assumption.  That is the assumption that 
only one effective operator dictates the result.  Our method is also model independent so any NP scenario involving LFV 
can be matched to the effective Lagrangian, Eq. \ref{Leff}.

In this work we assume that no new light particles (such as ``dark photons" or axions) exist in the low energy spectrum.
We do not consider neutrinos and we also assume that top quarks have been integrated out.  Charge-parity conservation is 
enforced, which restricts the WCs to the real number domain.  The effective operators are then written in terms of SM 
degrees of freedom such as leptons: $\ell_i = \tau, \mu,$ and $e$; and quarks: $b, c, s, u,$ and $d$.

Using these effective operators we consider the decays of heavy quark mesons and quarkonia to LFV states with and without photons: 
$\tau \mu \left(\gamma \right)$,  $\tau e \left(\gamma \right)$, and  $\mu e \left(\gamma \right)$.  Quarkonia are 
$q \overline{q}$ meson states.  For vector mesons we consider: $\Upsilon ( b \overline{b})$, $J/\psi (c \overline{c})$, 
$\phi (s \overline{s})$, $\rho \left(\frac{u \overline{u} - d \overline{d}}{\sqrt{2}} \right)$, 
$\omega \left(\frac{u \overline{u} + d \overline{d}}{\sqrt{2}} \right)$, and excited states.  For pseudo-scalar mesons we consider: 
$\eta_b (b \overline{b})$, $\eta_c (c \overline{c})$, $\eta \left(\frac{u \overline{u} + d \overline{d} -2s \overline{s}}{\sqrt{6}} \right)$, 
$\eta^{\prime} \left(\frac{u \overline{u} + d \overline{d} + s \overline{s}}{\sqrt{3}} \right)$, 
$\pi_0 \left(\frac{u \overline{u} - d \overline{d}}{\sqrt{2}} \right)$, $B_d^0 (d \overline{b})$, $B_s^0 (s \overline{b})$, $D^0 (c \overline{u})$, 
$K^0 (d \overline{s})$, and their excited states.  Finally for the scalar mesons we consider the quarkonia: 
$\chi_{b_0} (b \overline{b})$, $\chi_{c_0} (c \overline{c})$, and their excited states.

\begin{figure}[htbp]
    \centering
        \includegraphics[clip, trim=0cm 12.25cm 0cm 12.25cm, width=0.33\textwidth]{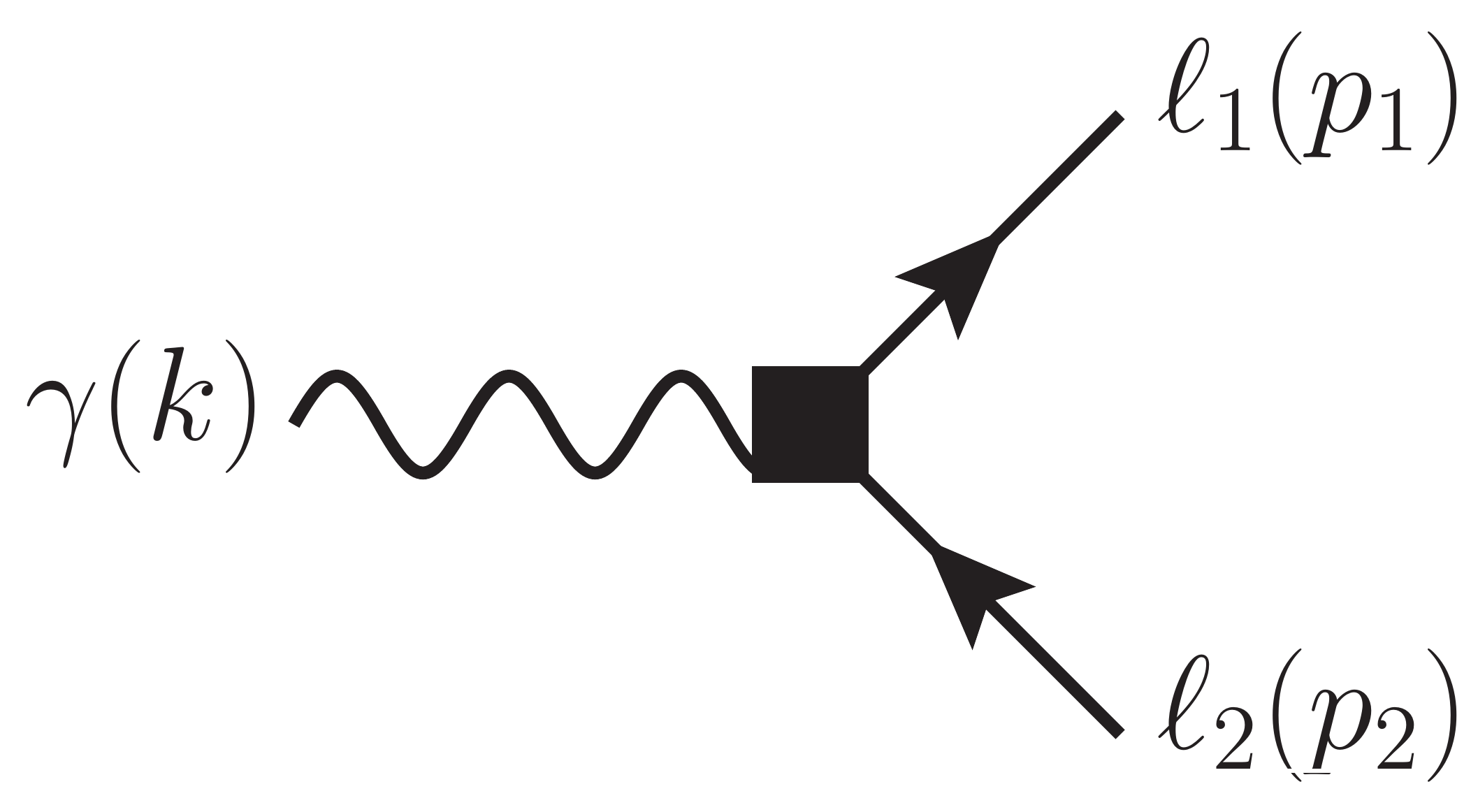}
        \vspace{2.5cm}
    \caption{Dipole interactions as described by ${\cal L}_D$, Eq. \ref{LD}.}
    \label{fig:LD}
\end{figure}

The effective Lagrangian, ${\cal L}_{\rm eff}$, can then be divided into a dipole part, ${\cal L}_D$; 
a part that involves four-fermion interactions, ${\cal L}_{\ell q}$; and a gluonic part, ${\cal L}_{G}$, \cite{Hazard:2016fnc}  
\begin{equation}\label{Leff}
{\cal L}_{\rm eff}= {\cal L}_D + {\cal L}_{\ell q} + {\cal L}_{G} + ... .
\end{equation}
Here the ellipses denote effective operators that are not relevant for the following analysis. 
The dipole part in Eq.~(\ref{Leff}) is usually written as \cite{Celis:2014asa}
\begin{eqnarray}\label{LD}
{\cal L}_{D} = -\frac{m_2}{\Lambda^2} \left[
\left( 
C_{DR}^{\ell_1\ell_2} \ \overline \ell_1 \sigma^{\mu\nu} P_L \ell_2 + 
C_{DL}^{\ell_1\ell_2} \ \overline \ell_1 \sigma^{\mu\nu} P_R \ell_2 
\right) F_{\mu\nu} + h.c. \right],
\end{eqnarray}
where $P_{\rm R,L}=(1\pm \gamma_5)/2$ is the right (left) chiral projection operator.  This dipole 
Lagrangian describes the interactions depicted in Fig. \ref{fig:LD}.  The WCs would, 
in general, be different for different leptons $\ell_i$. The constants $m_2$ and $\Lambda$ are the 
mass of the heavier lepton and the scale of new physics.  These dipole operators are selected by 
the quantum number of the two-body vector quarkonium decays.

The four-fermion dimension-six lepton-quark Lagrangian takes the form of
\begin{eqnarray}\label{Llq}
{\cal L}_{\ell q} = -\frac{1}{\Lambda^2} \sum_{q_1, q_2} \Big[
\left( C_{VR}^{q_1 q_2 \ell_1\ell_2} \ \overline\ell_1 \gamma^\mu P_R \ell_2 + 
C_{VL}^{q_1 q_2 \ell_1\ell_2} \ \overline\ell_1 \gamma^\mu P_L \ell_2 \right) \ \overline q_2 \gamma_\mu q_1 &&
\nonumber \\
+ \
\left( C_{AR}^{q_1 q_2 \ell_1\ell_2} \ \overline\ell_1 \gamma^\mu P_R \ell_2 + 
C_{AL}^{q_1 q_2 \ell_1\ell_2} \ \overline\ell_1 \gamma^\mu P_L \ell_2 \right) \ \overline q_2 \gamma_\mu \gamma_5 q_1 &&
\nonumber \\
+ \
m_2 m_{q_{\text{H}}} G_F \left( C_{SR}^{q_1 q_2 \ell_1\ell_2} \ \overline\ell_1 P_L \ell_2 + 
C_{SL}^{q\ell_1\ell_2} \ \overline\ell_1 P_R \ell_2 \right) \ \overline q_2 q_1 &&
\\
+ \
m_2 m_{q_{\text{H}}} G_F \left( C_{PR}^{q_1 q_2 \ell_1\ell_2} \ \overline\ell_1 P_L \ell_2 + 
C_{PL}^{q_1 q_2 \ell_1\ell_2} \ \overline\ell_1 P_R \ell_2 \right) \ \overline q_2 \gamma_5 q_1 
\nonumber \\
+ \
m_2 m_{q_{\text{H}}}G_F \left( C_{TR}^{q_1 q_2 \ell_1\ell_2} \ \overline\ell_1 \sigma^{\mu\nu} P_L \ell_2 + 
C_{TL}^{q_1 q_2 \ell_1\ell_2} \ \overline\ell_1 \sigma^{\mu\nu} P_R \ell_2 \right) \ \overline q_2 \sigma_{\mu\nu} q_1 
 &+& h.c. ~ \Big] .
\nonumber
\end{eqnarray}

\begin{figure}[htbp]
    \centering
        \includegraphics[clip, trim=0cm 0cm 1cm 0cm, width=0.375\textwidth]{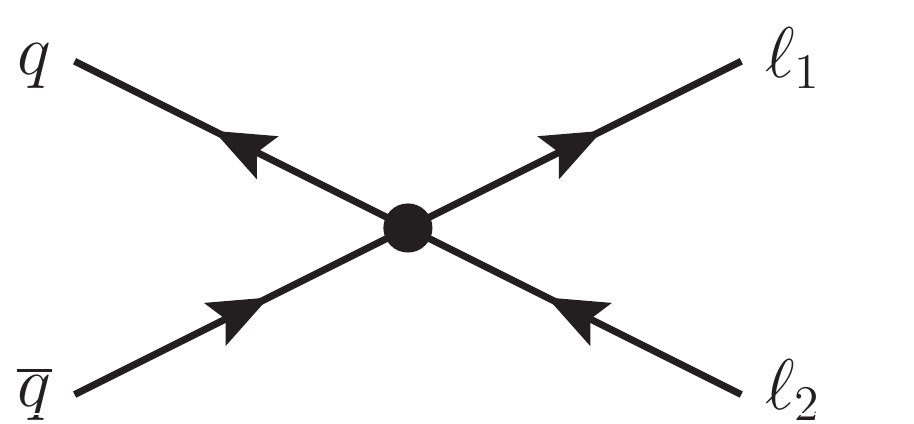}
        \vspace{0cm}
    \caption{Four-fermion interactions as described by ${\cal L}_{\ell q}$, Eq. \ref{Llq}.}
    \label{fig:Llq}
\end{figure}

This Lagrangian governs the four-fermion interactions depicted in Fig. \ref{fig:Llq} and depends on vector, 
axial-vector, scalar, pseudo-scalar, and tensor operators.  Each of these operators have associated WCs which, in 
general, are different for different quarks $q_i$ and leptons $\ell_j$.  Here $m_{q_{\text{H}}}$ is 
the mass of the heavier quark ($m_{q_{\text{H}}} = Max[m_{q_1},  m_{q_2}]$) and $G_F$ is Fermi's constant.  
The vector and tensor operators are selected by the quantum numbers of the two-body vector quarkonium decays.  
Likewise, the two body decays of pseudo-scalar mesons select the axial and pseudo-scalar operators, while the 
two-body decays of scalar quarkonia select the scalar operators.

The dimension seven gluonic operators can be either generated by some high scale physics or 
from integrating out heavy quark degrees of freedom \cite{Celis:2014asa,Petrov:2013vka},
\begin{eqnarray}\label{LG}
{\cal L}_{G} = -\frac{m_2 G_F}{\Lambda^2} \frac{\beta_L}{4\alpha_s} \Big[
\Big( C_{GR}^{\ell_1\ell_2} \ \overline\ell_1 P_L \ell_2 + 
C_{GL}^{\ell_1\ell_2} \ \overline\ell_1 P_R \ell_2 \Big)  G_{\mu\nu}^a G^{a \mu\nu} &&
\nonumber \\
+ ~ \Big( C_{\widetilde G R}^{\ell_1\ell_2} \ \overline\ell_1 P_L \ell_2 + 
C_{\widetilde G L}^{\ell_1\ell_2} \ \overline\ell_1 P_R \ell_2 \Big)  G_{\mu\nu}^a \widetilde G^{a \mu\nu}
 &+& h.c. \Big].
\end{eqnarray}
Here $\beta_L=-9 \alpha_s^2/(2\pi)$ is defined for the number of light active flavors, $L$, relevant to the scale 
of the process, which we take $\mu \approx 2$~GeV. All WCs 
should also be calculated at the same scale. The constant, $\alpha_s$, is the strong coupling constant and 
$\widetilde G^{a \mu\nu} = (1/2) \epsilon^{\mu\nu\alpha\beta} G^a_{\alpha\beta}$ is the dual to the
gluon field strength tensor \cite{Celis:2014asa}.

\begin{figure}[htbp]
    \centering
        \includegraphics[clip, trim=0cm 0cm 1cm 0cm, width=0.375\textwidth]{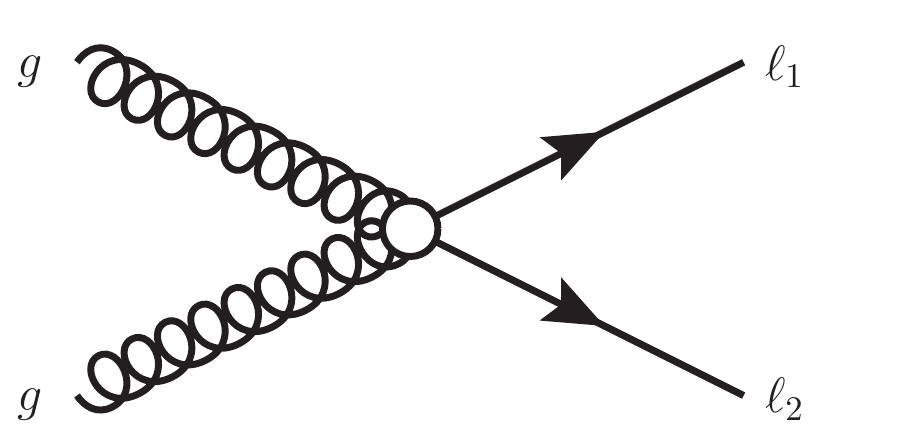}
        \vspace{0cm}
    \caption{Gluon-fermion interactions as described by ${\cal L}_{G}$, Eq. \ref{LG}.}
    \label{fig:LG}
\end{figure}

The gluonic Lagrangian, Eq. \ref{LG}, governs the gluon-fermion interactions depicted in Fig. \ref{fig:LG}. 
The gluonic operators associated with the gluon field strength tensor are selected by the two-body 
decays of scalar quarkonia, while the operators associated with it dual are selected by the two-body 
decays of pseudo-scalar quarkonia.

\section{Two-body vector quarkonium decays $V \to \ell_1 \overline \ell_2$} \label{sec:2bodyVec}

\begin{figure}[htbp]
    \centering
        \includegraphics[clip, trim=4.2cm 5cm 0cm 11.5cm, width=0.66\textwidth]{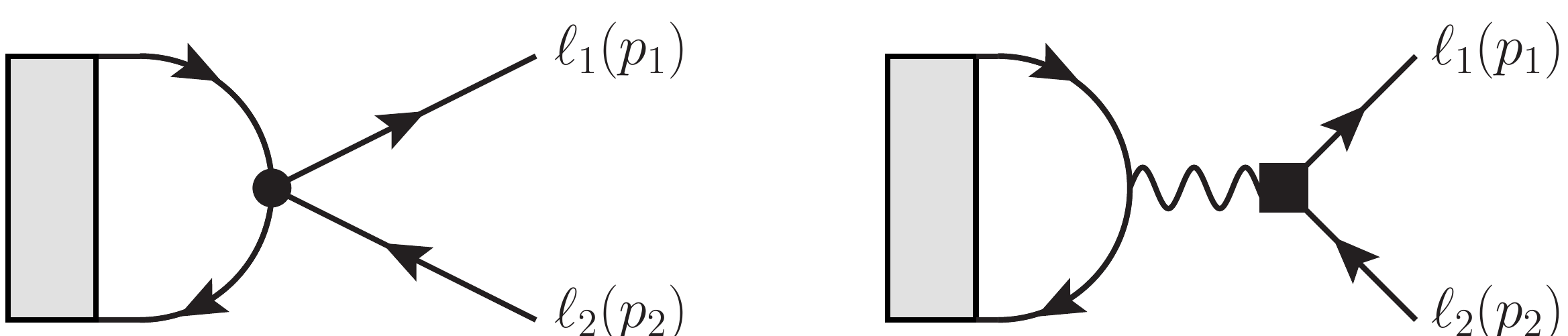}
        \vspace{0cm}
    \caption{Feynmann diagrams for $V \to \ell_1 \overline \ell_2$ decay.}
    \label{fig:VecAmp}
\end{figure}

Vector quarkonia are mesons of type $q \overline{q}$ with quantum numbers $J^{PC}=1^{--}$ such as 
$\Upsilon$, $J/\psi$, $\phi$, $\rho$, and $\omega$.  Here we denote them with the letter $V$. The amplitude 
for these decays receives contributions from two Feynman diagrams shown in Fig. \ref{fig:VecAmp} and takes 
the general form of Eq. \ref{eqn:VecAmp}.  It depends on the vector and tensor decay constants defined in 
Eqs. \ref{eqn:VecDecayConst} and \ref{eqn:TenDecayConst} \cite{Becirevic:2013bsa}.  These constants can be derived either from 
experimental data or calculated using methods of lattice quantum chromodynamics (QCD).  In Fig. \ref{fig:VecAmp} the circular black vertex 
represents the contributions from the vector and tensor operators from the four-fermion Lagrangian, Eq. \ref{Llq}, 
and the black square represents the dipole operator contribution from Eq. \ref{LD}.

\begin{eqnarray}
\langle 0| \overline q \gamma^\mu q | V(p) \rangle &=& f_V m_V \epsilon^\mu (p) \label{eqn:VecDecayConst} \\
\langle 0| \overline q \sigma^{\mu\nu} q | V(p) \rangle &=& i f^T_V \left( \epsilon^\mu p^\nu-p^\mu \epsilon^\nu\right)
\label{eqn:TenDecayConst}
\end{eqnarray}

The most general amplitude for $V \to \ell_1 \overline{\ell}_2$ transitions can be written

\begin{eqnarray}\label{eqn:VecAmp}
{\cal A}(V\to \ell_1 \overline \ell_2) = \overline{u}(p_1, s_1) \left[
A_V^{\ell_1\ell_2} \gamma_\mu + B_V^{\ell_1\ell_2} \gamma_\mu \gamma_5 
+ \frac{C_V^{\ell_1\ell_2}}{m_{V}} (p_2-p_1)_\mu 
\right. ~~~~~~~~~~~~~~
\nonumber \\
\qquad + \left.
\frac{iD_V^{\ell_1\ell_2}}{m_{V} }(p_2-p_1)_\mu \gamma_5 \
\right] v(p_2,s_2) \ \epsilon^\mu(p) \text{.}
\end{eqnarray}

This amplitude is dependent on four dimensionless constants $A_V^{\ell_1\ell_2}$, $B_V^{\ell_1\ell_2}$, 
$C_V^{\ell_1\ell_2}$, and $D_V^{\ell_1\ell_2}$, which contain the WCs and decay constants.  From 
this general amplitude we are able to calculate the branching ratio for $V \to \ell_1 \overline \ell_2$ in 
Eq. \ref{eqn:VecBR}, which depends on the dimensionless constants, Eqs.  \ref{eqn:VCoefA}--\ref{eqn:VCoefD}.

\begin{eqnarray}
\tfrac{{\cal B}(V \to \ell_1 \overline \ell_2)}{{\cal B}(V \to e^+e^-)} \! \! &=& \! \!
\left(\tfrac{m_V \left(1-y^2\right)}{4\pi\alpha  f_V Q_q}\right)^2 \! \Big[ \! \left(\left|A_V^{\ell_1\ell_2}\right|^2 \! \! +  
\left|B_V^{\ell_1\ell_2}\right|^2\right) \! + \! \tfrac{1}{2} \! \left(1 \! - \! 2y^2\right) \!  
\left(\left|C_V^{\ell_1\ell_2}\right|^2 \! \! + \left|D_V^{\ell_1\ell_2}\right|^2\right)  \nonumber \\
&& + y \, \text{Re} \! \left(A_V^{\ell_1\ell_2} C_V^{\ell_1\ell_2 *}+i B_V^{\ell_1\ell_2} D_V^{\ell_1\ell_2 *}\right) 
\Big] \label{eqn:VecBR}
\end{eqnarray}

\begin{eqnarray}
\! \! \! \! \! \! \! \! \! \! A_V^{\ell_1\ell_2} \! \! \! \! &=&\! \! \tfrac{f_V m_V}{\Lambda^2} \! \left[ \ \
\sqrt{4\pi \alpha} Q_{q}  y^2 (C_{DL}^{\ell_1\ell_2}+C_{DR}^{\ell_1\ell_2}) + 
\kappa_V (C_{VL}^{q \ell_1\ell_2} + C_{VR}^{q \ell_1\ell_2}) \right. \label{eqn:VCoefA} \\
&& \qquad\qquad + \left. 2 y^2 \kappa_V \frac{f^T_V}{f_V} G_F m_V m_{q}   
(C_{TL}^{q \ell_1\ell_2} + C_{TR}^{q \ell_1\ell_2})
 \right]
\nonumber \\
\! \! \! \! \! \! \! \! \! \! B_V^{\ell_1\ell_2} \! \! \! \! &=& \! \! \tfrac{f_V m_V}{\Lambda^2} \! \left[
- \sqrt{4\pi \alpha} Q_{q}   y^2 (C_{DL}^{\ell_1\ell_2}-C_{DR}^{\ell_1\ell_2}) - 
\kappa_V (C_{VL}^{q \ell_1\ell_2} - C_{VR}^{q \ell_1\ell_2}) \right.  \label{eqn:VCoefB} \\
&& \qquad\qquad - \left. 2 y^2 \kappa_V \frac{f^T_V}{f_V} G_F m_V m_{q}   
(C_{TL}^{q \ell_1\ell_2} -C_{TR}^{q \ell_1\ell_2})
 \right] \nonumber \\
\! \! \! \! \! \! \! \! \! \! C_V^{\ell_1\ell_2} \! \! \! \! &=& \! \! \tfrac{ f_V m_V}{\Lambda^2} y \left[ \ \ \sqrt{4\pi \alpha} Q_{q} 
(C_{DL}^{\ell_1\ell_2} + C_{DR}^{\ell_1\ell_2}) + 
2 \kappa_V \tfrac{f^T_V}{f_V}
G_F m_V m_{q} (C_{TL}^{q \ell_1\ell_2} + C_{TR}^{q \ell_1\ell_2}) \right] 
\label{eqn:VCoefC} \\
\! \! \! \! \! \! \! \! \! \! D_V^{\ell_1\ell_2} \! \! \! \! &=& \! \! \! \! i \tfrac{ f_V m_V}{\Lambda^2} y \left[ -\sqrt{4\pi \alpha} Q_{q} 
(C_{DL}^{\ell_1\ell_2} - C_{DR}^{\ell_1\ell_2}) -
2 \kappa_V \tfrac{f^T_V}{f_V}
G_F m_V m_{q}  (C_{TL}^{q \ell_1\ell_2} - C_{TR}^{q \ell_1\ell_2}) \right] 
\label{eqn:VCoefD}
\end{eqnarray}

Note in Eqs. \ref{eqn:VecBR}--\ref{eqn:VCoefD} that $m_V$ is the mass of the vector meson, $y$ is the ratio of the heavier lepton mass to the vector meson mass ($\tfrac{m_2}{m_V}$), $Q_{q}$ is the charge of the quark ($\tfrac{2}{3}$, $-\tfrac{1}{3}$), $\alpha$ is the fine structure constant, and $\kappa_V$ is a state ($V$) dependent constant.  For pure $q \overline{q}$ states $\kappa_V = \tfrac{1}{2}$. We also suppress all quark subindices (i.e. $q_{1,2} \to q$) and abbreviate the WC indices (i.e. $C_{VR(L)}^{q_1 q_2 \ell_1 \ell_2} \to C_{VR(L)}^{q \ell_1 \ell_2}$) in the case of quarkonia because $q_1 = q_2$. One will also note that there is indeed dipole, vector, and tensor operator dependence for the vector decays.  One can see this because of the appearance of the appropriate WCs in the dimensionless constants  $A_V^{\ell_1\ell_2}$-- $D_V^{\ell_1\ell_2}$.  

It is also important to note that the decay constants approximately cancel when the branching ratio is normalized to the usually well known branching ratio of $V \to e^+ e^-$.  We are left with the ratio of $\tfrac{f^T_V}{f_V}$ for the tensor operator contributions to Eqs. \ref{eqn:VCoefA}--\ref{eqn:VCoefD}.  The tensor decay constants, $f^T_V$, are not well known with the exception of $f^T_{J/\psi}$, which is equal to $ 410 \pm 10$ MeV \cite{Becirevic:2013bsa}.  The decay constants used in these calculations can be found in Table. \ref{tab:Vdecay_constants}.  The values of tensor decay constants are expected to be in the same range.  Examining the decay constant values for $J/\psi$ \cite{Becirevic:2013bsa} we see that their central values are within a few percent of each other.  Knowing this, we make the necessary assumption in our calculations that $f_V \approx f^T_V$ (as suggested in Ref. \cite{Khodjamirian:2015dda}).

\begin{table}
\begin{center}
\footnotesize
\begin{tabular}{|c|c|c|c|c|c|c|c|}
\hline\hline
~State & $~\Upsilon(1S)~$ & $~\Upsilon(2S)~$ & $~\Upsilon(3S)~$ & $~J/\psi~$ & $~\psi(2S)$~  & $~\phi~$ & $~\rho \left(\omega \right)~$\\
\hline
~$f_V$, MeV~ &  $649\pm 31$   &  $481\pm 39$ &  $539\pm 84$ &  $418\pm 9$ &  $294\pm 5$ &  $241\pm 18$ & $209.4\pm 1.5$\\
\hline\hline
\end{tabular}
\normalsize
\end{center}
\caption{Vector meson decay constants used in the calculation of branching ratios ${\cal B}(V \to \ell_1 \overline \ell_2)$. 
The transverse decay constants are set $f^T_V=f_V$ except for $J/\psi$, which has 
$f^T_{J/\psi} = 410 \pm 10$ MeV
\cite{Becirevic:2013bsa,Abada:2015zea,Colquhoun:2014ica,MaiordeSousa:2012vv,Donald:2013pea,Chen:2015tpa}.
}\label{tab:Vdecay_constants} 
\end{table} 

To constrain the WCs and thus new physics we need to know the experimental upper limits on the branching ratios of LFV decays, which can be found in Table. \ref{tab:Vdecaylimits}.  We know that the SM predicts the LFV decay of $\mu \to e \gamma$ to be of the order $\sim 10^{-54}$ \cite{Cheng:1985bj} and current experimental techniques cannot hope to reach that level of sensitivity.  If there is NP, then it must be above this threshold, but below current experimental limits.

\begin{table*}
\begin{center}
\footnotesize
\begin{tabular}{lccc}
\hline \hline
$\ell_1 \ell_2$ &$\mu \tau$ & $e \tau$ & $e \mu$  \\ 
\hline
${\cal B}(\Upsilon (1S) \to \ell_1 \ell_2)$ & $ 6.0 \times 10^{-6}$ & $-$ & $-$  \\
${\cal B}(\Upsilon (2S) \to \ell_1 \ell_2)$ &  $3.3 \times 10^{-6}$ & $3.2 \times 10^{-6}$ & $-$ \\ 
${\cal B}(\Upsilon (3S) \to \ell_1 \ell_2)$ &  $3.1 \times 10^{-6}$ & $4.2 \times 10^{-6}$ & $-$ \\
${\cal B}(J/\psi \to \ell_1 \ell_2)$ &  $2.0 \times 10^{-6}$ & $8.3 \times 10^{-6}$ & $1.6 \times 10^{-7}$ \\
${\cal B}(\phi \to \ell_1 \ell_2)$ &  n/a & n/a & $4.1 \times 10^{-6}$ \\ 
${\cal B}(\ell_2 \to \ell_1 \gamma)$ & $4.4 \times 10^{-8}$ & $3.3 \times 10^{-8}$ & $5.7 \times 10^{-13}$ \\
\hline \hline
\end{tabular}
\end{center}
\caption{\label{tab:Vdecaylimits} Available experimental upper bounds on 
${\cal B}(V \to \ell_1 \ell_2)$ and ${\cal B}( \ell_2 \to \ell_1 \gamma)$ \cite{PDG,Lees:2010jk}. 
Dashes signify that no experimental constraints are available and ``n/a" means 
that the transition is forbidden by available phase space. Charge averages of the final states are always assumed.}
\end{table*}

For the limits in Table. \ref{tab:Vdecaylimits} we are able to constrain the WCs of the vector and 
tensor operators in Table. \ref{tab:V4fermion}.  The keen observer will notice that the limit on the 
tensor WCs for the $s$-quark does not make sense.  We expect that the WCs should 
be of order one and $\Lambda$ is some unknown higher scale of NP.  The constrained ratio of 
$|C_{TR(L)}^{s \ell_1 \ell_2}|/\Lambda^2$ should be less than one GeV$^{-2}$.  This is common in effective field theory (EFT) analyses 
and is not an indication of a breakdown of the EFT, it simply means that the experimental constraints are 
not strong enough to give us a meaningful answer (see, e.g. \cite{Petrov:2013nia}).

\begin{table*}
\begin{center}
\footnotesize
\begin{tabular}{ccccccc}
\hline \hline
Wilson coeff. & Leptons &\multicolumn{5}{c}{Initial state (quark)}\\
(GeV$^{-2}$) & $\ell_1 \ell_2$ & $\Upsilon(1S) \ (b)$ & $\Upsilon(2S) \ (b)$ & $\Upsilon(3S) \ (b)$ 
 & $J/\psi \ (c)$ & $\phi \ (s)$  \\ \hline
$~$ & $\mu \tau$ & $ 5.6 \times 10^{-6}$ & $4.1 \times 10^{-6}$ & $3.5 \times 10^{-6}$ 
 & $5.5 \times 10^{-5}$ & n/a \\
$\left| C_{VL}^{q\ell_1\ell_2}/\Lambda^2 \right|$ & $e \tau$ & $-$ & $4.1 \times 10^{-6}$ & $4.1 \times 10^{-6}$ 
 & $1.1 \times 10^{-4}$ & n/a \\ 
$~$ & $e \mu$ & $-$ & $-$ & $-$  
& $1.0 \times 10^{-5}$ & $2 \times 10^{-3}$  \\
\hline
$~$ & $\mu \tau$  & $ 5.6 \times 10^{-6}$ & $4.1 \times 10^{-6}$ & $3.5 \times 10^{-6}$ 
 & $5.5 \times 10^{-5}$ & n/a \\
$\left| C_{VR}^{q\ell_1\ell_2}/{\Lambda^2} \right|$ & $e \tau$ & $-$ & $4.1 \times 10^{-6}$ & $4.1 \times 10^{-6}$ 
 & $1.1 \times 10^{-4}$ & n/a \\
$~$ & $e \mu$ & $-$ & $-$ & $-$ 
 & $1.0 \times 10^{-5}$ & $2 \times 10^{-3}$ \\
\hline
$~$  & $\mu \tau$  & $ 4.4 \times 10^{-2}$ & $3.2 \times 10^{-2}$ & $2.8 \times 10^{-2}$ 
 & $1.2$ & n/a \\
$\left| {C_{TL}^{q\ell_1\ell_2}}/{\Lambda^2} \right|$ & $e \tau$ & $-$ & $3.3 \times 10^{-2}$ & $3.2 \times 10^{-2}$ 
 & $2.4$ & n/a \\
$~$ & $e \mu$ & $-$ & $-$ & $-$ 
 & $4.8$ & $1 \times 10^{4}$ \\
\hline
$~$  & $\mu \tau$  & $ 4.4 \times 10^{-2}$ & $3.2 \times 10^{-2}$ & $2.8 \times 10^{-2}$ 
 & $1.2$ & n/a \\
$\left| {C_{TR}^{q\ell_1\ell_2}}/{\Lambda^2} \right|$ & $e \tau$ & $-$ & $3.3 \times 10^{-2}$ & $3.2 \times 10^{-2}$ 
 & $2.4$ & n/a \\
$~$ & $e \mu$ & $-$ & $-$ & $-$ 
 & $4.8$ & $1 \times 10^{4}$ \\
 \hline \hline
\end{tabular}
\end{center}
\caption{\label{tab:V4fermion}Constraints on the WCs of four-fermion operators. Dashes signify that 
no experimental data are available to produce a constraint; ``n/a" means that the transition is forbidden by phase space.
Note that no experimental data is available for higher excitations of $\psi$ (from \cite{Hazard:2016fnc}).}
\end{table*}

We can also constrain the dipole WCs as seen in Table. \ref{tab:Vdipoles}.  Radiative LFV decays of 
charged leptons give much stronger constrains on these operators, but the vector meson decay 
can still provide complimentary constraints.

\begin{table*}
\begin{center}
\footnotesize
\begin{tabular}{cccccccc}
\hline \hline
Wilson coeff. & Leptons &\multicolumn{5}{c}{Initial state} & \\
(GeV$^{-2}$) & $\ell_1 \ell_2$ & $\Upsilon(1S) \ (b)$ & $\Upsilon(2S) \ (b)$ & $\Upsilon(3S) \ (b)$ 
 & $J/\psi \ (c)$ & $\phi (s)$ & $\ell_2 \to \ell_1 \gamma$ \\ \hline
$~$ & $\mu \tau$ & $2.0 \times 10^{-4}$ & $1.6 \times 10^{-4}$ & $1.4 \times 10^{-4}$ 
 & $2.5 \times 10^{-4}$ & n/a & $2.6 \times 10^{-10}$ \\
$\left| C_{DL}^{\ell_1\ell_2}/\Lambda^2 \right|$ & $e \tau$ & $-$ & $1.6 \times 10^{-4}$ & $1.6 \times 10^{-4}$ 
 & $5.3 \times 10^{-4}$ & n/a & $2.7 \times 10^{-10}$ \\
$~$ & $e \mu$ & $-$ & $-$ & $-$  
& $1.1 \times 10^{-3}$ & $0.2$ & $3.1 \times 10^{-7}$ \\
\hline
$~$ & $\mu \tau$  & $2.0 \times 10^{-4}$ & $1.6 \times 10^{-4}$ & $1.4 \times 10^{-4}$ 
 & $2.5 \times 10^{-4}$ & n/a & $2.6 \times 10^{-10}$ \\
$\left| C_{DR}^{\ell_1\ell_2}/{\Lambda^2} \right|$ & $e \tau$ & $-$ & $1.6 \times 10^{-4}$ & $1.6 \times 10^{-4}$ 
 & $5.3 \times 10^{-4}$ & n/a & $2.7 \times 10^{-10}$ \\
$~$ & $e \mu$ & $-$ & $-$ & $-$ 
 & $1.1 \times 10^{-3}$ & $0.2$ & $3.1 \times 10^{-7}$ \\
\hline \hline
\end{tabular}
\end{center}
\caption{\label{tab:Vdipoles} Constraints on the dipole WCs from the $1^{--}$ quarkonium
decays and radiative lepton transitions $\ell_2 \to \ell_1 \gamma$. Dashes signify that 
no experimental data are available to produce a constraint; ``n/a" means that the transition is 
forbidden by phase space (from \cite{Hazard:2016fnc}).}
\end{table*}

\section{Two-body pseudo-scalar meson decays $P \to \ell_1 \overline \ell_2$}

\begin{figure}[htbp]
    \centering
        \includegraphics[clip, trim=0cm 11.5cm 0cm 11.5cm, width=0.66\textwidth]{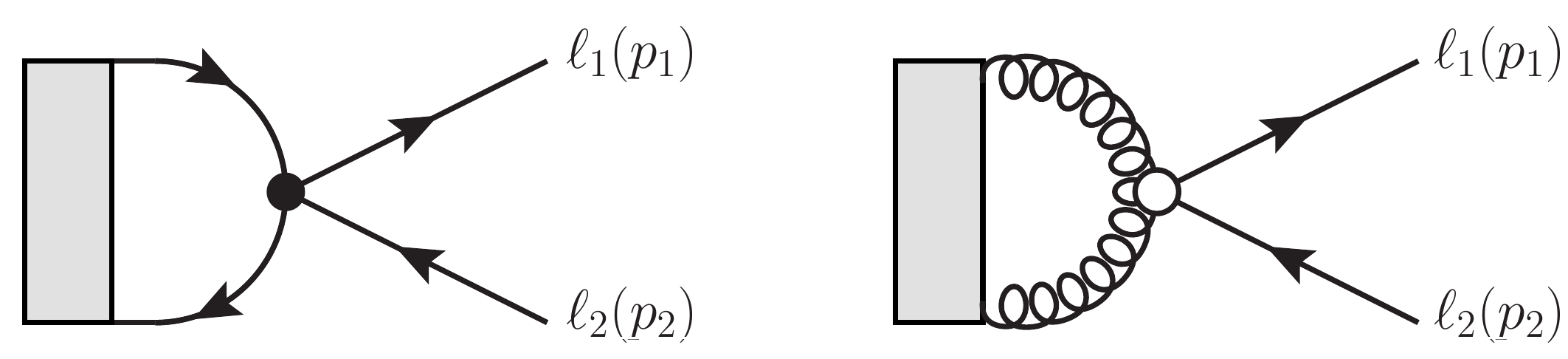}
        \vspace{0cm}
    \caption{Feynmann diagrams for $P(S) \to \ell_1 \overline \ell_2$ decay.}
    \label{fig:PAmp}
\end{figure}

Similar to the vector meson decays, we can look at the decays of pseudo-scalar mesons.  Pseudo-scalar 
mesons have quantum numbers $J^{PC}=0^{-+}$.  Examples include $\eta_b$, $\eta_c$, $\eta$, $\eta'$, 
$\pi^0$, $B_{d(s)}^0$, $D^0$, and $K^0$.  The amplitude for pseudo-scalar decays includes contributions 
from the diagrams in Fig. \ref{fig:PAmp}.  Here the solid and open black circles are effective vertices that 
depend on the axial and pseudo-scalar operator contributions of ${\cal L}_{\ell q}$, Eq. \ref{Llq}, and 
the gluonic operators in Eq. \ref{LG}.

\begin{eqnarray}
&& \langle 0| \overline{q}_2 \gamma^\mu \gamma_5 q_1 | P(p) \rangle = -i f_P p^\mu\,, 
\label{eqn:PseDecayConstant} \\
&& \langle 0| \frac{\alpha_s}{4\pi} G^{a\mu\nu} \widetilde G^a_{\mu\nu}  | P(p) \rangle = a_P \,.
\label{eqn:PseAnomElement}
\end{eqnarray}

The generic amplitude for these pseudo-scalar meson decays is shown in Eq. \ref{eqn:PAmp}, which depends on two dimensionless constants $E_P^{\ell_1\ell_2}$ and $F_P^{\ell_1\ell_2}$.  These constants depend on the WCs of the axial, pseudo-scalar, and gluonic effective operators and the decay constants defined in Eqs. \ref{eqn:PseDecayConstant} and \ref{eqn:PseAnomElement} \cite{Petrov:2013vka}.

\begin{eqnarray}\label{eqn:PAmp}
{\cal A}(P\to \ell_1 \overline \ell_2) = \overline{u}(p_1, s_1) \left[
E_P^{\ell_1\ell_2}  + i F_P^{\ell_1\ell_2} \gamma_5 
\right] v(p_2,s_2) \,
\end{eqnarray}

This generic amplitude yields the branching ratio for $P \to \ell_1 \overline{\ell}_2$ in Eq. \ref{eqn:SPBR}.  This branching ratio depends on the dimensionless constants found in Eqs. \ref{eqn:SPCoefE} and \ref{eqn:SPCoefF}.  The coefficient equations presented here are only valid for pure $q_1 \overline{q}_2$ states.  For mixed states the situation is similar, but more involved.  For details see \cite{Hazard:2016fnc} and \cite{HazardPetrovFuture}.

\begin{eqnarray}\label{eqn:SPBR}
{\cal B}(P \to \ell_1 \overline \ell_2) = \frac{m_P}{8\pi \Gamma_P} \left(1-y^2\right)^2
\left[\left|E_P^{\ell_1\ell_2}\right|^2 + \left|F_P^{\ell_1\ell_2}\right|^2\right]
\end{eqnarray}

\begin{eqnarray}
E_P^{\ell_1\ell_2} &=& \,\,\,y \frac{m_P}{4 \Lambda^2} \Big[
- i f_{P} \left( 2 \left(C_{AL}^{q_1 q_2 \ell_1\ell_2}+C_{AR}^{q_1 q_2 \ell_1\ell_2}\right) - m_{P}^{2} G_{F} \left( C_{PL}^{q_1 q_2 \ell_1\ell_2} + C_{PR}^{q_1 q_2 \ell_1\ell_2} \right) \right) \nonumber \\
&& \hspace{2.75in}  + 9 G_F a_{P} \left(C_{\widetilde G L}^{\ell_1\ell_2} + C_{\widetilde G R}^{\ell_1\ell_2} \right) \Big], \label{eqn:SPCoefE} \\
F_P^{\ell_1\ell_2} &=& \!\!\!\!  -iy \frac{m_P}{4 \Lambda^2} \Big[-if_{P}\left( 2 \left(C_{AL}^{q_1 q_2 \ell_1\ell_2}-C_{AR}^{q_1 q_2 \ell_1\ell_2}\right) - m_P^2 G_F \left(C_{PL}^{q_1 q_2 \ell_1\ell_2}-C_{PR}^{q_1 q_2 \ell_1\ell_2}\right) \right) \nonumber \\
&& \hspace{2.75in} + 9 G_F a_{P} \left(C_{\widetilde G L}^{\ell_1\ell_2} - C_{\widetilde G R}^{\ell_1\ell_2} \right) \Big].  \label{eqn:SPCoefF}
\end{eqnarray}

In Eqs. \ref{eqn:SPBR} -- \ref{eqn:SPCoefF} the constant $m_P$ is the pseudo-scalar meson mass, $y= \tfrac{m_2}{m_P}$, and $\Gamma_P$ is the pseudo-scalar meson total decay rate.

To constrain the WCs we need to know the decay constants (Tables. \ref{tab:Pdecay_constants} and \ref{tab:Pdecayconstants}), the anomalous matrix element values, and the experimental limits (Tables. \ref{tab:Pdecaylimitsqq} and \ref{tab:Pdecaylimitsq1q2}) for the pseudo-scalar mesons.  The anomalous matrix elements do not contribute for non-quarkonium states (i.e. $q_1 \neq q_2$).  For the quarkonium states we have $a_{\eta} = -0.022 \pm 0.002$ GeV$^3$ and $a_{\eta^\prime} = -0.057 \pm 0.002$ GeV$^3$ \cite{Beneke:2002jn}.  For heavy quarks $q=c,b$ one expects 
the anomalous matrix elements to be quite small and so we take $a_{\eta_{b(c)}} \approx 0$ \cite{Hazard:2016fnc}.

\begin{table}
\begin{center}
\footnotesize
\begin{tabular}{c|ccccccc}
\hline\hline
~State & $~\eta_b~$ & $~\eta_c~$ & $\eta, u(d)$ & $~\eta, s$~  & $~\eta^\prime, u(d)~$  & $~\eta^\prime, s~$ & $~\pi~$\\
~$f_P^q$, MeV~ &  $667 \pm 6$   &  $387\pm 7$ &  $108\pm 3$ &  $-111\pm 6$ &  $89\pm 3$ &   $136\pm 6$&   $130.41 \pm 0.20$ \\
\hline\hline
\end{tabular}
\normalsize
\end{center}
\caption{Pseudoscalar meson decay constants used in the calculation of branching ratios ${\cal B}(P \to \ell_1 \overline \ell_2)$ for operators of type ${\cal O} \sim q \overline{q} \ell_1 \overline{\ell}_2$ \cite{Becirevic:2013bsa,PDG,McNeile:2012qf,Beneke:2002jn}.}
\label{tab:Pdecay_constants} 
\end{table} 

\begin{table*}
\begin{center}
\footnotesize
\begin{tabular}{c|cccc}
\hline \hline
~State & $B^0_d$ & $B^0_s$ & $D^0$ & $K^0_L$ \\
~$f_P$, MeV~ &  $186 \pm 4$   &  $224 \pm 4$ &  $207.4 \pm 3.8$ &  $155.0 \pm 1.9$ \\
\hline \hline
\end{tabular}
\end{center}
\caption{\label{tab:Pdecayconstants} Pseudoscalar meson decay constants used in the calculation of branching ratios ${\cal B}(P \to \ell_1 \overline \ell_2)$ for operators of type ${\cal O} \sim q_1 \overline{q}_2 \ell_1 \overline{\ell}_2$ ($q_1 \neq q_2$) \cite{Dowdall:2013tga,Carrasco:2014poa}.}
\end{table*} 

\begin{table*}
\begin{center}
\footnotesize
\begin{tabular}{cc}
\hline \hline
$\ell_1 \ell_2$ & $e \mu$  \\ 
\hline
${\cal B}(\eta \to \ell_1 \ell_2)$ & $6 \times 10^{-6}$ \\ 
${\cal B}(\eta^{\prime} \to \ell_1 \ell_2)$ & $4.7 \times 10^{-4}$ \\
${\cal B}(\pi^0 \to \ell_1 \ell_2)$ & $3.6 \times 10^{-10}$ \\
\hline \hline
\end{tabular}
\end{center}
\caption{\label{tab:Pdecaylimitsqq} Available experimental limits on ${\cal B}(P \to \ell_1 \ell_2)$ for operators of type ${\cal O} \sim q \overline{q} \ell_1 \overline{\ell}_2$ \cite{PDG}. Charge averages of the final states are always assumed.}
\end{table*}

\begin{table*}
\begin{center}
\footnotesize
\begin{tabular}{cccc}
\hline \hline
$\ell_1 \ell_2$& $\mu \tau$ & $e \tau$ & $e \mu$  \\ 
\hline
${\cal B}(B^0_d \to \ell_1 \ell_2)$ & $2.2 \times 10^{-5}$ & $2.8 \times 10^{-5}$ & $2.8 \times 10^{-9}$ \\ 
${\cal B}(B^0_s \to \ell_1 \ell_2)$ & $-$ & $-$ & $1.1 \times 10^{-8}$ \\
${\cal B}(D^0 \to \ell_1 \ell_2)$ &  n/a & $-$ & $2.6 \times 10^{-7}$ \\
${\cal B}(K^0_L \to \ell_1 \ell_2)$ &  n/a &  n/a & $4.7 \times 10^{-12}$ \\
\hline \hline
\end{tabular}
\end{center}
\caption{\label{tab:Pdecaylimitsq1q2} Available experimental limits on ${\cal B}(P \to \ell_1 \ell_2)$ for operators of type ${\cal O} \sim q_1 \overline{q}_2 \ell_1 \overline{\ell}_2$ ($q_1 \neq q_2$) \cite{PDG}.
Dashes signify that no experimental data are available; ``n/a" means that the transition is forbidden by phase space. Charge averages of the final states are always assumed.}
\end{table*}

Given the limits in Tables. \ref{tab:Pdecaylimitsqq} and \ref{tab:Pdecaylimitsq1q2} we are able to constrain the WCs for the axial, pseudo-scalar, and gluonic operators as shown in Tables. \ref{tab:P4fermionqq}--\ref{tab:P4fermionq1q2}.  Again we see that the constraints on the pseudo-scalar and gluonic WCs over $\Lambda^2$ are a two or more orders of magnitude greater than one.  This is not the result of a breakdown of the EFT.  It occurs because the experimental constraints are not strong enough.  

\begin{table*}
\begin{center}
\footnotesize
\begin{tabular}{cccccccc}
\hline \hline
 Wilson coefficient & Leptons &\multicolumn{6}{c}{Initial state}\\
 (GeV$^2$) & $\ell_1 \ell_2$ & $\eta_b$ & $\eta_c$ & $\eta (u/d)$ & $\eta (s)$ & $\eta^\prime (u/d)$ & $\eta^\prime (s)$\\ \hline
$~$ & $\mu \tau$ & $-$ & $-$ & n/a & n/a & n/a & n/a\\
$\left| {C_{AL}^{q\ell_1\ell_2}}/{\Lambda^2} \right|$ & $e \tau$ & $-$ & $-$ & n/a & n/a & n/a & n/a \\
$~$ & $e \mu$ & $-$ & $-$ & $3 \times 10^{-3}$ & $2 \times 10^{-3}$ & $2.1 \times 10^{-1}$ & $1.9 \times 10^{-1}$\\
\hline
$~$ & $\mu \tau$ & $-$ & $-$ & n/a & n/a & n/a & n/a\\
$\left| {C_{AR}^{q\ell_1\ell_2}}/{\Lambda^2} \right|$ & $e \tau$ & $-$ & $-$ & n/a & n/a & n/a & n/a\\
$~$ & $e \mu$ & $-$ & $-$ & $3 \times 10^{-3}$ & $2 \times 10^{-3}$ & $2.1 \times 10^{-1}$ & $1.9 \times 10^{-1}$\\
\hline
$~$ & $\mu \tau$ & $-$ & $-$ & n/a & n/a & n/a & n/a\\
$\left| {C_{PL}^{q\ell_1\ell_2}}/{\Lambda^2} \right|$ & $e \tau$ & $-$ & $-$ & n/a & n/a & n/a & n/a\\
$~$ & $e \mu$ & $-$ & $-$ & $2 \times 10^{3}$ & $1 \times 10^{3}$ & $3.9 \times 10^{4}$ & $3.6 \times 10^{4}$\\
\hline
$~$ & $\mu \tau$ & $-$ & $-$ & n/a & n/a & n/a & n/a\\
$\left| {C_{PR}^{q\ell_1\ell_2}}/{\Lambda^2} \right|$ & $e \tau$ & $-$ & $-$ & n/a & n/a & n/a & n/a\\
$~$ & $e \mu$ & $-$ & $-$ & $2 \times 10^{3}$ & $1 \times 10^{3}$ & $3.9 \times 10^{4}$ & $3.6 \times 10^{4}$\\
\hline \hline
\end{tabular}
\end{center}
\caption{\label{tab:P4fermionqq}Constraints on the WCs from pseudo-scalar quarkonium decays. Dashes signify that 
no experimental data is available to produce a constraint; ``n/a" means that the transition is forbidden by phase space (from \cite{Hazard:2016fnc}). Note that the WC indices are abbreviated for quarkonia (i.e. $C_{AR(L)}^{q_1 q_2 \ell_1 \ell_2} \to C_{AR(L)}^{q \ell_1 \ell_2}$).}
\end{table*}

\begin{table*}
\begin{center}
\footnotesize
\begin{tabular}{cccccc}
\hline \hline
Wilson coefficient & Leptons &\multicolumn{4}{c}{Initial state} \\
(GeV$^{-2}$) & $\ell_1 \ell_2$ & $\eta_b$ & $\eta_c$ & $\eta$ & $\eta^{\prime}$ \\
\hline
$\left| C_{GL}^{\ell_1\ell_2}/\Lambda^2 \right|$  & $e \mu$ & $-$ & $-$ & $2 \times 10^2$  & $5.0 \times 10^{3}$ \vspace{0.1cm} \\
$\left| C_{GR}^{\ell_1\ell_2}/{\Lambda^2} \right|$ & $e \mu$ & $-$ & $-$ & $2 \times 10^2$ & $5.0 \times 10^{3}$ \vspace{0.1cm} \\
\hline \hline
\end{tabular}
\end{center}
\caption{\label{tab:Pgluon} Constraints on the pseudo-scalar gluonic WCs. Dashes signify that 
no experimental data is available to produce a constraint. No data for other lepton species is available (from \cite{Hazard:2016fnc}).}
\end{table*}

\begin{table*}
\begin{center}
\footnotesize
\begin{tabular}{cccccc}
\hline \hline
 & Leptons &\multicolumn{4}{c}{Initial state}\\
 Wilson coefficient & $\ell_1 \ell_2$ & $B^0_d \left(d \bar b\right)$ & $B^0_s\left(s \bar b\right)$ & $D^0 \left(c \bar u\right)$ & $K^0_L \left(\frac{d \bar s - s \bar d}{\sqrt{2}}\right)$ \\ \hline
$~$ & $\mu \tau$ & $2.3 \times 10^{-8}$ & $-$ & n/a & n/a \\
$\left| {C_{AL}^{q_1 q_2 \ell_1\ell_2}}/{\Lambda^2} \right|$ & $e \tau$ & $2.6 \times 10^{-8}$ & $-$ & $-$ & n/a  \\
$~$ & $e \mu$ & $3.9 \times 10^{-9}$ & $6.3 \times 10^{-9}$ & $1.1 \times 10^{-7}$ & $5.0 \times 10^{-12}$ \\
\hline
$~$ & $\mu \tau$ & $2.3 \times 10^{-8}$ & $-$ & n/a & n/a \\
$\left| {C_{AR}^{q_1 q_2 \ell_1\ell_2}}/{\Lambda^2} \right|$ & $e \tau$ & $2.6 \times 10^{-8}$ & $-$ & $-$ & n/a \\
$~$ & $e \mu$ & $3.9 \times 10^{-9}$ & $6.3 \times 10^{-9}$ & $1.1 \times 10^{-7}$ & $5.0 \times 10^{-12}$ \\
\hline
$~$ & $\mu \tau$ & $7.1 \times 10^{-5}$ & $-$ & n/a & n/a \\
$\left| {C_{PL}^{q_1 q_2 \ell_1\ell_2}}/{\Lambda^2} \right|$ & $e \tau$ & $8.0 \times 10^{-5}$ & $-$ & $-$ & n/a \\
$~$ & $e \mu$ & $1.2 \times 10^{-5}$ & $1.9 \times 10^{-5}$ & $2.7 \times 10^{-3}$ & $1.7 \times 10^{-6}$ \\
\hline
$~$ & $\mu \tau$ & $7.1 \times 10^{-5}$ & $-$ & n/a & n/a \\
$\left| {C_{PR}^{q_1 q_2 \ell_1\ell_2}}/{\Lambda^2} \right|$ & $e \tau$ & $8.0 \times 10^{-5}$ & $-$ & $-$ & n/a \\
$~$ & $e \mu$ & $1.2 \times 10^{-5}$ & $1.9 \times 10^{-5}$ & $2.7 \times 10^{-3}$ & $1.7 \times 10^{-6}$ \\
\hline \hline
\end{tabular}
\end{center}
\caption{\label{tab:P4fermionq1q2}Constraints on the WCs from pseudo-scalar meson decays. Center dots signify that 
no experimental data are available to produce a constraint; ``n/a" means that the transition is forbidden by phase space (from \cite{HazardPetrovFuture}).}
\end{table*}

\section{Two-body scalar meson decays $S \to \ell_1 \overline \ell_2$}

We can perform this exercise a third time for scalar quarkonia, which would give ideal access to the scalar and gluonic effective operators in Eqs. \ref{Llq} and \ref{LG}.  These are quarkonia with quantum numbers $0^{++}$ like $\chi_{b0}$ and $\chi_{c0}$.  The amplitude for scalar decays includes contributions from the Feynman diagrams in Fig. \ref{fig:PAmp}.  The solid black circular vertex represents the scalar effective operator contibutions while the open black circle represents the gluonic contributions.
\begin{eqnarray}\label{eqn:}
&&  \langle 0| \overline q q | S(p) \rangle = -i m_S f_S  \label{eqn:SDecayConstant} \\
&& \langle 0| \frac{\alpha_s}{4\pi} G^{a\mu\nu} G^a_{\mu\nu}  | S(p) \rangle = a_S \label{eqn:SAnomEmement}
\end{eqnarray}

The amplitude will be dependent on a decay constant and an anomalous matrix element, which are defined in Eqs. \ref{eqn:SDecayConstant} and \ref{eqn:SAnomEmement} \cite{Godfrey:2015vda}.  The most general expression for the $S \to \ell_1 \overline \ell_2$ decay amplitude looks exactly like 
Eq.~(\ref{eqn:PAmp}), with obvious modifications for the scalar decay:
\begin{eqnarray}\label{eqn:SAmp}
{\cal A}(S\to \ell_1 \overline \ell_2) = \overline{u}(p_1, s_1) \left[
E_S^{\ell_1\ell_2}  + i F_S^{\ell_1\ell_2} \gamma_5 
\right] v(p_2,s_2) .
\end{eqnarray}
$E_S^{\ell_1\ell_2}$ and $F_S^{\ell_1\ell_2}$ are dimensionless constants. The branching ratio, which follows from 
Eq.~(\ref{eqn:SAmp}), is
\begin{eqnarray}\label{BRSpin0S}
{\cal B}(S \to \ell_1 \overline \ell_2) = \frac{m_S}{8\pi \Gamma_S} \left(1-y^2\right)^2
\left[\left|E_S^{\ell_1\ell_2}\right|^2 + \left|F_S^{\ell_1\ell_2}\right|^2\right].
\end{eqnarray}
Here $\Gamma_S$ is the total width of the scalar state, $m_S$ is the meson mass, and $y = m_2/m_S$. 
The coefficients $E_S^{\ell_1\ell_2}$ and $F_S^{\ell_1\ell_2}$  are
\begin{eqnarray}\label{SCoef1}
E_S^{\ell_1\ell_2} &=& y \frac{m_S G_F}{4 \Lambda^2} 
\left[2 i f_{S} m_S m_q \left(C_{SL}^{q l_1 l_2} + C_{SR}^{q l_1 l_2}\right) +
9 a_S \left(C_{GL}^{q l_1 l_2} + C_{GR}^{q l_1 l_2}\right) \right],
\nonumber \\
F_S^{\ell_1\ell_2} &=& y \frac{m_S G_F}{4 \Lambda^2}
 \left[2 f_{S} m_S m_q \left(C_{SL}^{q l_1 l_2} - C_{SR}^{q l_1 l_2}\right) -
9 i a_S \left(C_{GL}^{q l_1 l_2} - C_{GR}^{q l_1 l_2}\right)
\right].
\end{eqnarray}

Currently there are no experimental limits on the LFV decays of scalar mesons.  The state $\chi_{b0}$ and $\chi_{c0}$ could be produced via gluon-gluon fusion at the LHC, B decays at flavor factories, or radiative decays of $\Upsilon(2S)$, $\Upsilon(3S)$, $\psi(2S)$, or $\psi(3770)$.

\section{Resonant transitions of three-body vector quarkonium decays}

The radiative decays of vector mesons where there is a scalar meson resonance, $V \to \gamma \left(S \to \ell_1 \ell_2 \right)$, may be used to study the two-body decays of scalar mesons.  If the soft photon, $\gamma$, can be tagged at B-factories, then the branching ratios factorize into ${\cal B}\left(V \to \ell_1 \overline{\ell}_2 \right) = {\cal B}\left(V \to \gamma S \right) {\cal B}\left(S \to \ell_1 \overline{\ell}_2 \right)$.  This would be quite useful as the the relevant decays of $\psi$ and $\Upsilon$ are the order of $1-10 \%$ \cite{PDG}. In charm,
\begin{eqnarray}\label{BranchRadc}
&&  {\cal B}(\psi(2S) \to \gamma \chi_{c0} (1P)) = 9.99 \pm 0.27\% \ ,
\nonumber \\
&& {\cal B}(\psi(3770) \to \gamma \chi_{c0}(1P)) = 0.73 \pm 0.09\% \ .
\nonumber 
\end{eqnarray}
The corresponding radiative transitions in the beauty sector are also rather large,
\begin{eqnarray}\label{BranchRadb}
&& {\cal B}(\Upsilon(2S) \to \gamma \chi_{b0}(1P)) = 3.8 \pm 0.4\% \ ,
\nonumber \\
&& {\cal B}(\Upsilon(3S) \to \gamma \chi_{b0}(1P)) = 0.27 \pm 0.04\% \ ,
\\
&&  {\cal B}(\Upsilon(3S) \to \gamma \chi_{b0} (2P)) = 5.9 \pm 0.6\% \ .
\nonumber 
\end{eqnarray}

\section{Non-resonant transitions of three-body vector quarkonium decays}

\begin{figure}
\center
\subfigure[]{\includegraphics[scale=0.45]{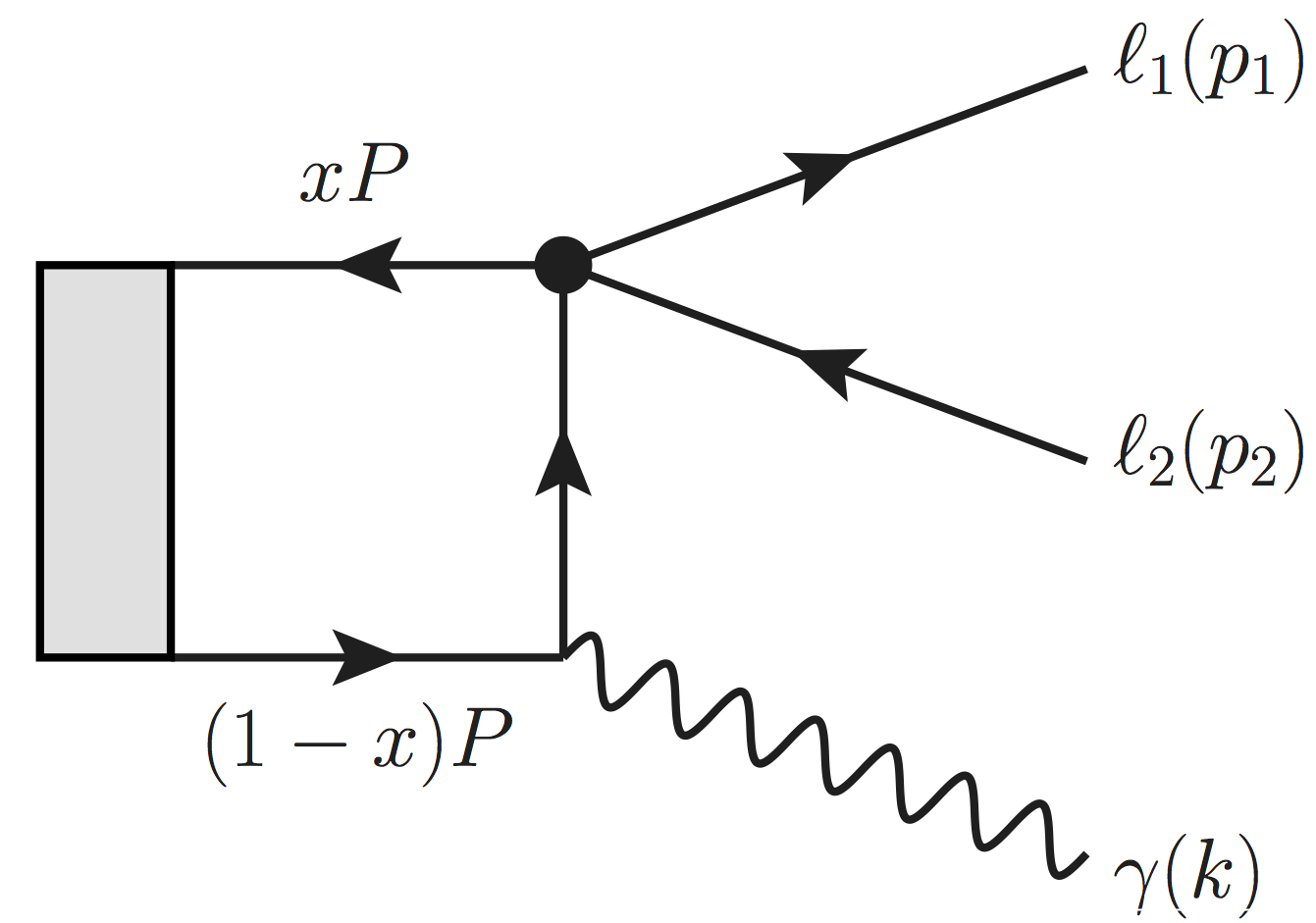} \label{diagram a}}
\subfigure[]{\includegraphics[scale=0.45]{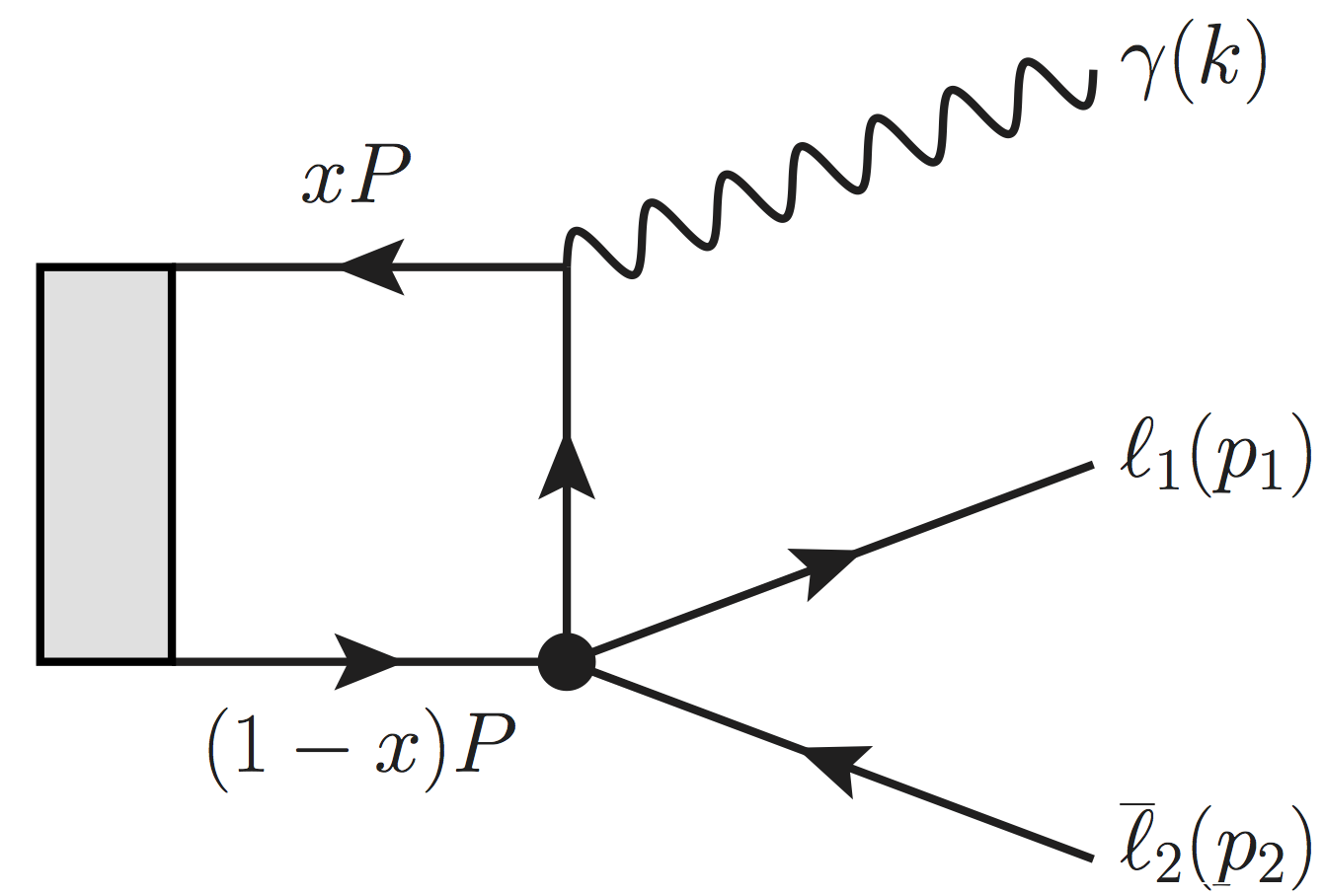} \label{diagram b}} \\
\subfigure[]{\includegraphics[scale=0.45]{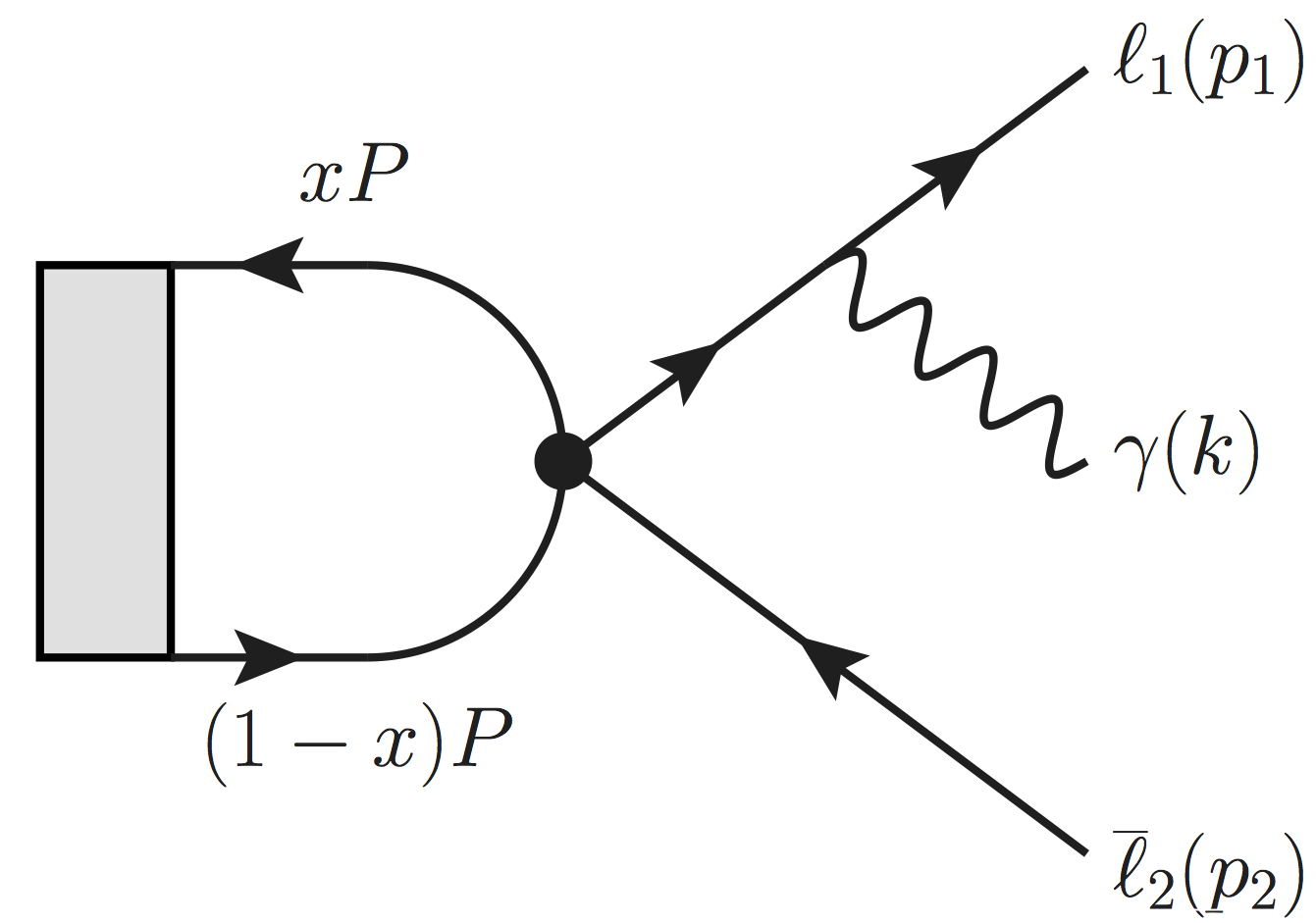} \label{diagram c}}
\subfigure[]{\includegraphics[scale=0.45]{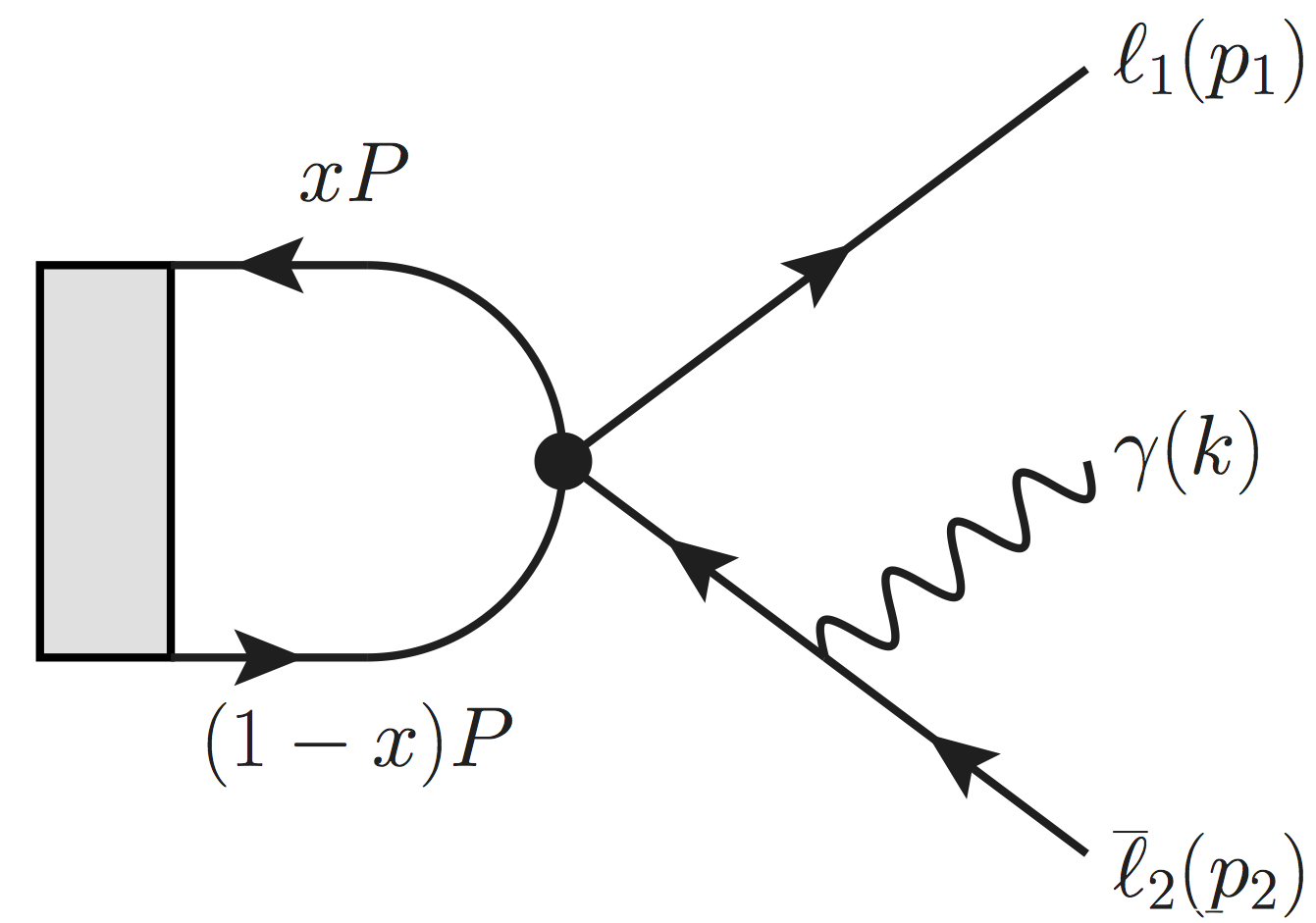} \label{diagram d}} \\
\subfigure[]{\includegraphics[scale=0.45]{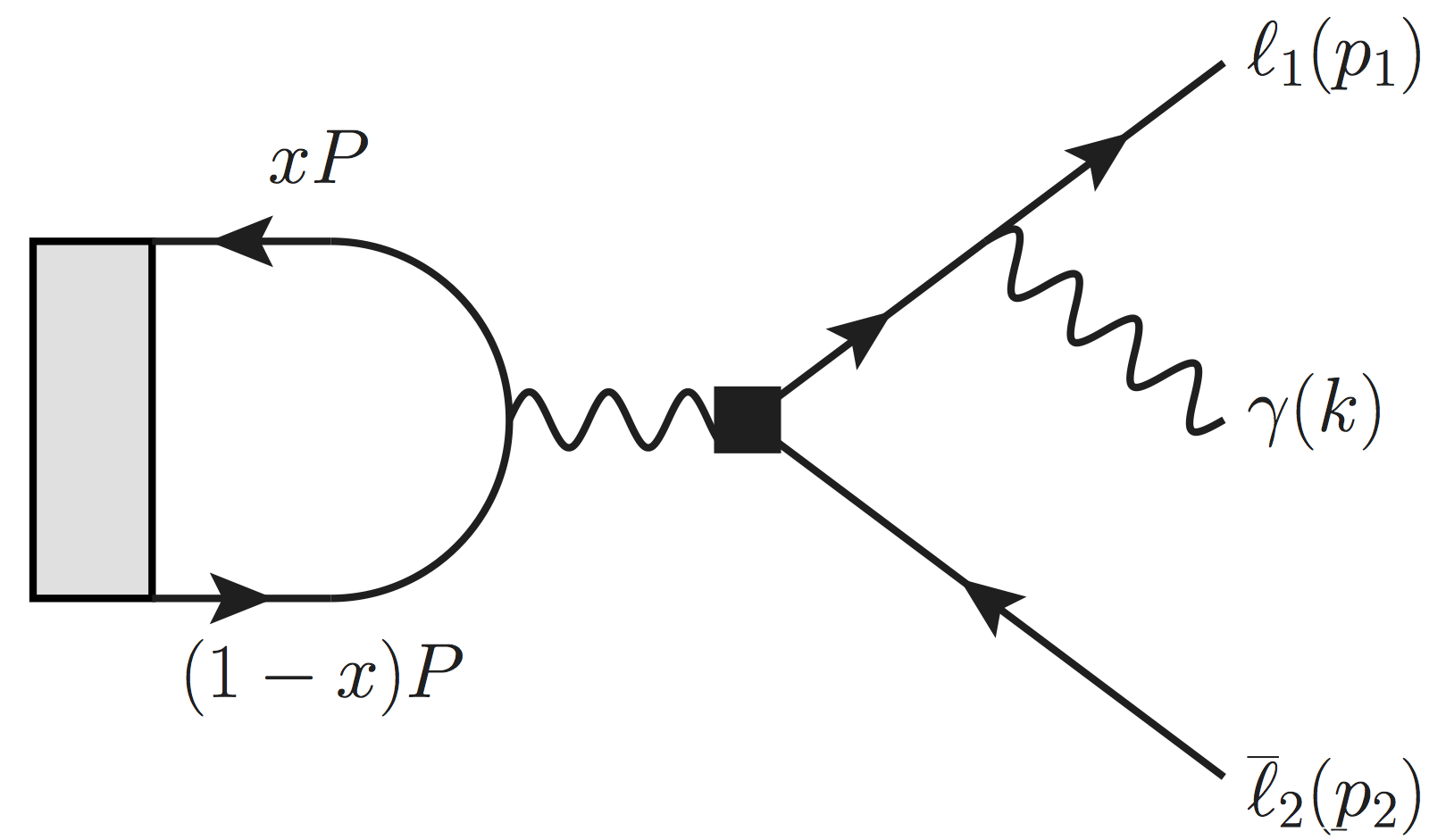} \label{diagram e}}
\subfigure[]{\includegraphics[scale=0.45]{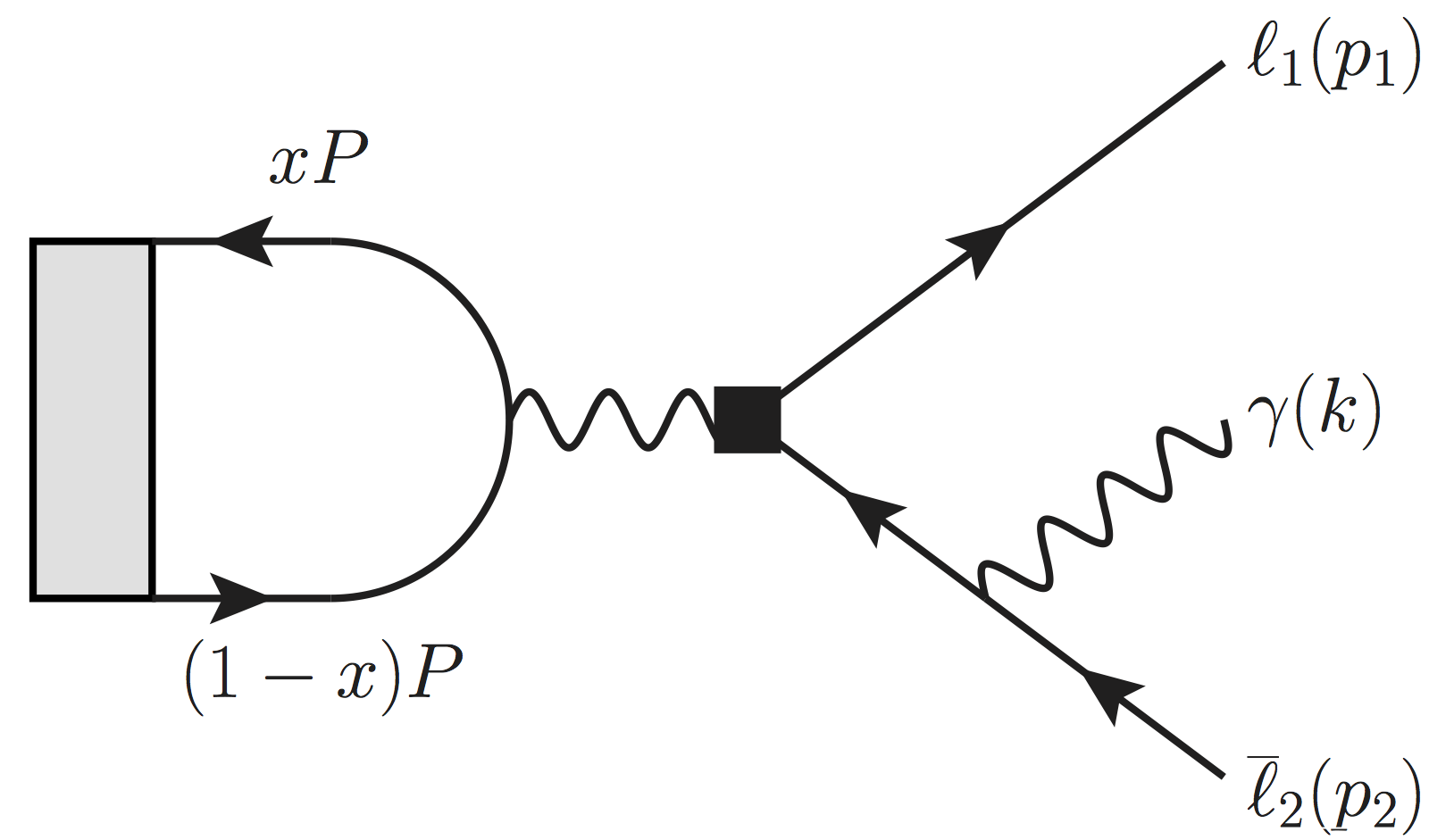} \label{diagram f}} \\
\subfigure[]{\includegraphics[scale=0.45]{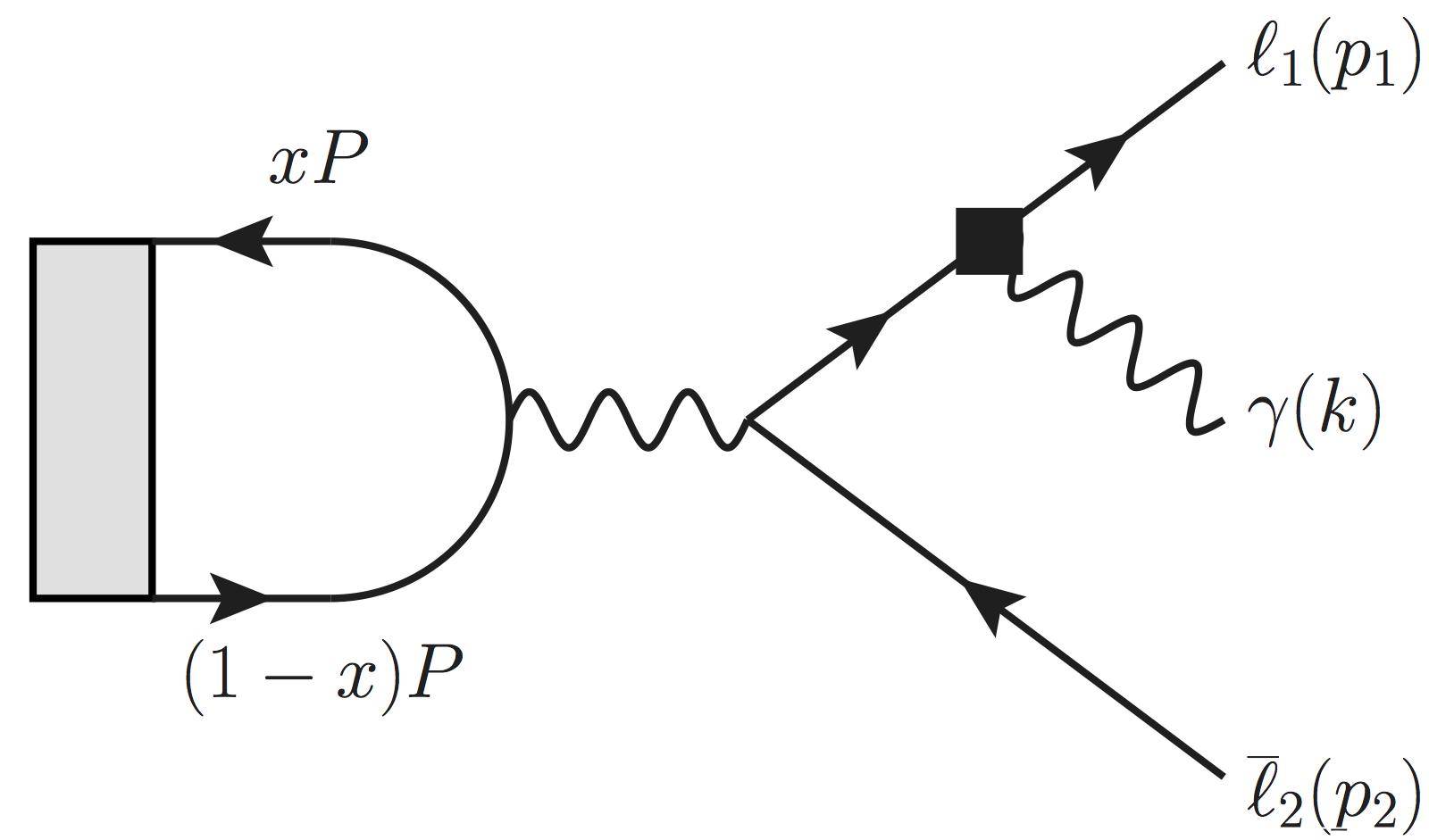} \label{diagram g}}
\subfigure[]{\includegraphics[scale=0.45]{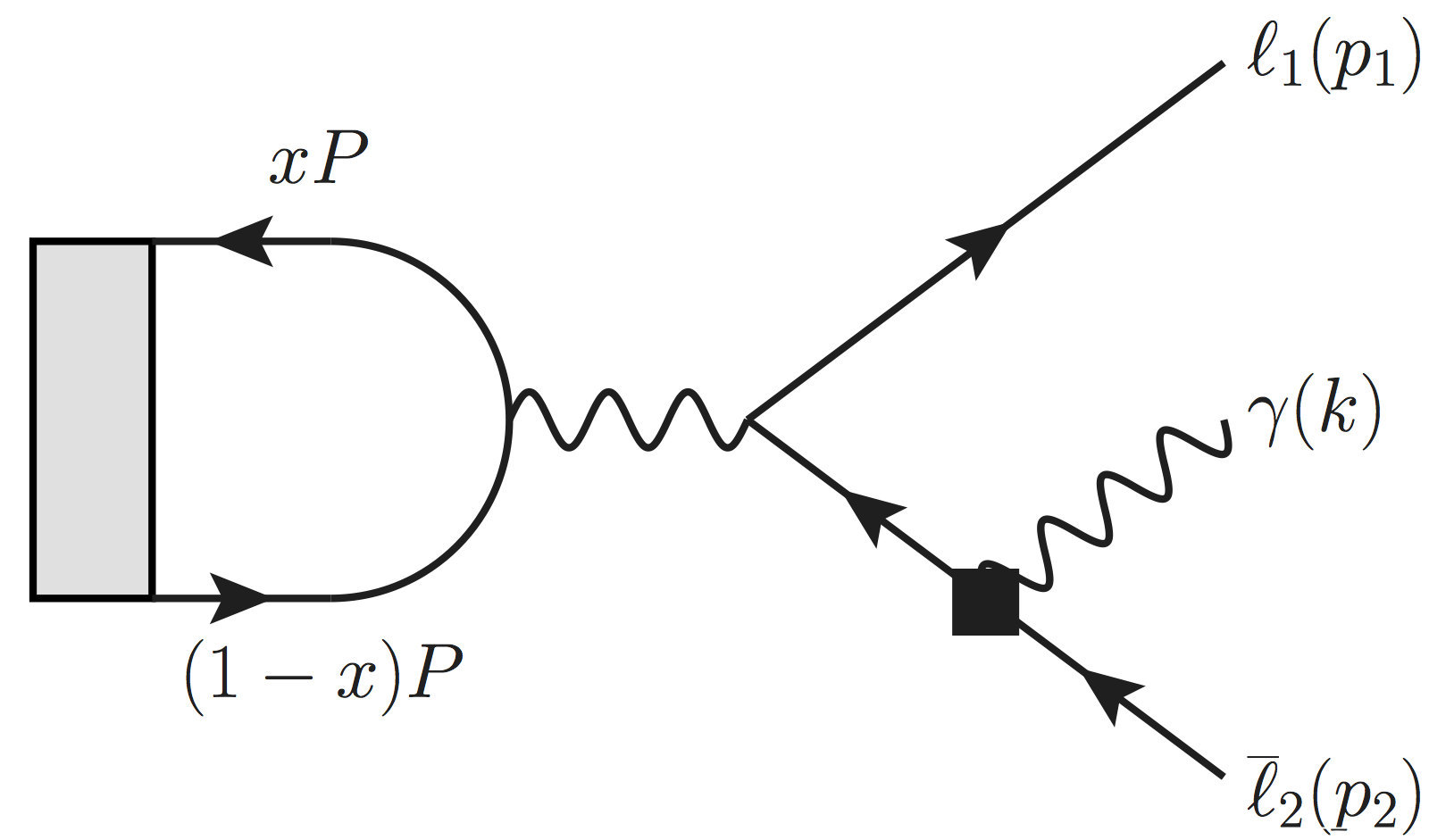} \label{diagram h}}
\caption{ Feynman diagrams for ${\cal A} (V \to \gamma \ell_1 \overline \ell_2)$. 
The black circles represent the four fermion LFV vertex, the black boxes represent the 
dipole LFV vertex, and the grey boxes represent the quarkonium bound state.}
\label{fig:vector3body}
\end{figure}

Non-resonant three-body decays of vector quarkonia, $V \to \gamma \ell_1 \overline{\ell}_2$, 
could be used to constrain the scalar operators in Eq. \ref{Llq}.  This calculation would involve 
all eight of the diagrams in Fig. \ref{fig:vector3body}, but only Figs. \ref{diagram a} and \ref{diagram b} 
contain the scalar operators.  The black circles represent the four-fermion LFV vertex 
interactions (Eq. \ref{Llq}) while the black boxes represent the dipole vertex (Eq. \ref{LD}). 
Ideally we would perform this calculation in a model independent manner using $V \to \gamma$ 
form-factors.  Unfortunately the necessary form-factors are not well known and, at the cost of 
our model independence, we turn to a constituent-quark model to access $C_{SL(R)}^{q \ell_1 \ell_2}$.

For our calculation we choose a model with a quark-antiquark wave function of the vector quarkonium state \cite{Aditya:2012ay, Dziembowski:1986dr, Szczepaniak:1990dt, Lepage:1980fj}
\begin{eqnarray}
\Psi_V = \frac{I_c}{\sqrt{6}} \Phi_V(x) \left(m_V \gamma^{\alpha}+ i p^{\beta} \sigma^{\alpha \beta} \right) \epsilon^{\alpha}(p).
\label{wavefunction}
\end{eqnarray} 

Where $I_c$ is the color space identity matrix; $m_V$, $p$, and $\varepsilon(p)$ are the vector quarkonium mass, momentum, and polarization vector; $x$ is the quarkonium momentum fraction carried by one of the quarks, and $\Phi_V(x)$ is the distribution amplitude defined as

\begin{eqnarray}\label{distributionfunction}
\Phi_V(x) = \frac{f_V}{2\sqrt{6}}  \delta(x-1/2).
\end{eqnarray}

This simple wave function was chosen to approximate each quark carrying half the meson's momentum.  The nonlocal matrix element is then calculated using 

\begin{eqnarray}
\braket{0|\overline q \Gamma^{\mu} q}V =\int_{0}^{1} \text{Tr}[\Gamma^{\mu} \Psi_V] dx  \label{matrixelement}.
\end{eqnarray}.

Using the nonlocal matrix element and assuming \textit{single operator dominance} we are able to calculate the differential decay rates for the axial, scalar, and pseudo-scalar operators from Figs. \ref{diagram a} and \ref{diagram b}:

\begin{eqnarray} \label{3bodydiffdecayrates}
\frac{d\Gamma_{V \to \gamma \ell_1 \overline \ell_2}^A}{dm_{12}^2} &=& \frac{1}{9} \frac{\alpha Q_q^2}{\left(4 \pi\right)^2} 
\frac{f_V^2}{\Lambda^4} \left(C_{AL}^2+C_{AR}^2 \right) \frac{\left(m_V^2-m_{12}^2\right) 
\left(2 m_V^2 y^2 + m_{12}^2\right) \left(m_V^2 y^2 - m_{12}^2\right)^2}{m_V m_{12}^6}, 
\nonumber \\
\frac{d\Gamma_{V \to \gamma \ell_1 \overline \ell_2}^S}{dm_{12}^2} &=& \frac{1}{24} 
\frac{\alpha Q_q^2}{\left(4 \pi\right)^2} \frac{f_V^2 G_F^2 m_V}{\Lambda^4} 
\left(C_{SL}^2+C_{SR}^2 \right) \frac{y^2 \left(m_V^2-m_{12}^2\right) 
\left(m_V^2 y^2 - m_{12}^2\right)^2}{m_{12}^2}, 
\\
\frac{d\Gamma_{V \to \gamma \ell_1 \overline \ell_2}^P}{dm_{12}^2} &=& \frac{1}{24} 
\frac{\alpha Q_q^2}{\left(4 \pi\right)^2} \frac{f_V^2 G_F^2 m_V}{\Lambda^4} 
\left(C_{PL}^2+C_{PR}^2 \right) \frac{y^2 \left(m_V^2-m_{12}^2\right) 
\left(m_V^2 y^2 - m_{12}^2\right)^2}{m_{12}^2}.
\nonumber
\end{eqnarray}

\begin{figure}
\subfigure[]{\includegraphics[scale=.63]{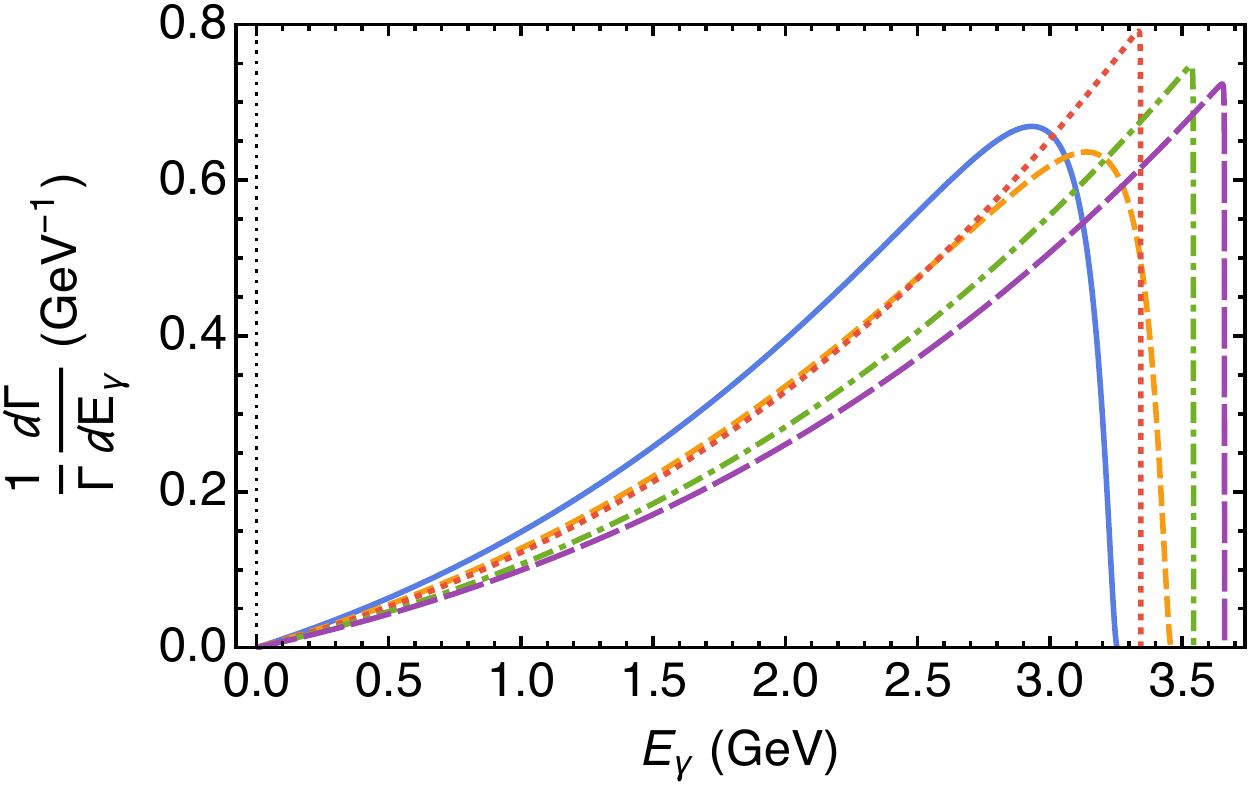} \label{diffdecayaxialbb}}
\subfigure[]{\includegraphics[scale=.63]{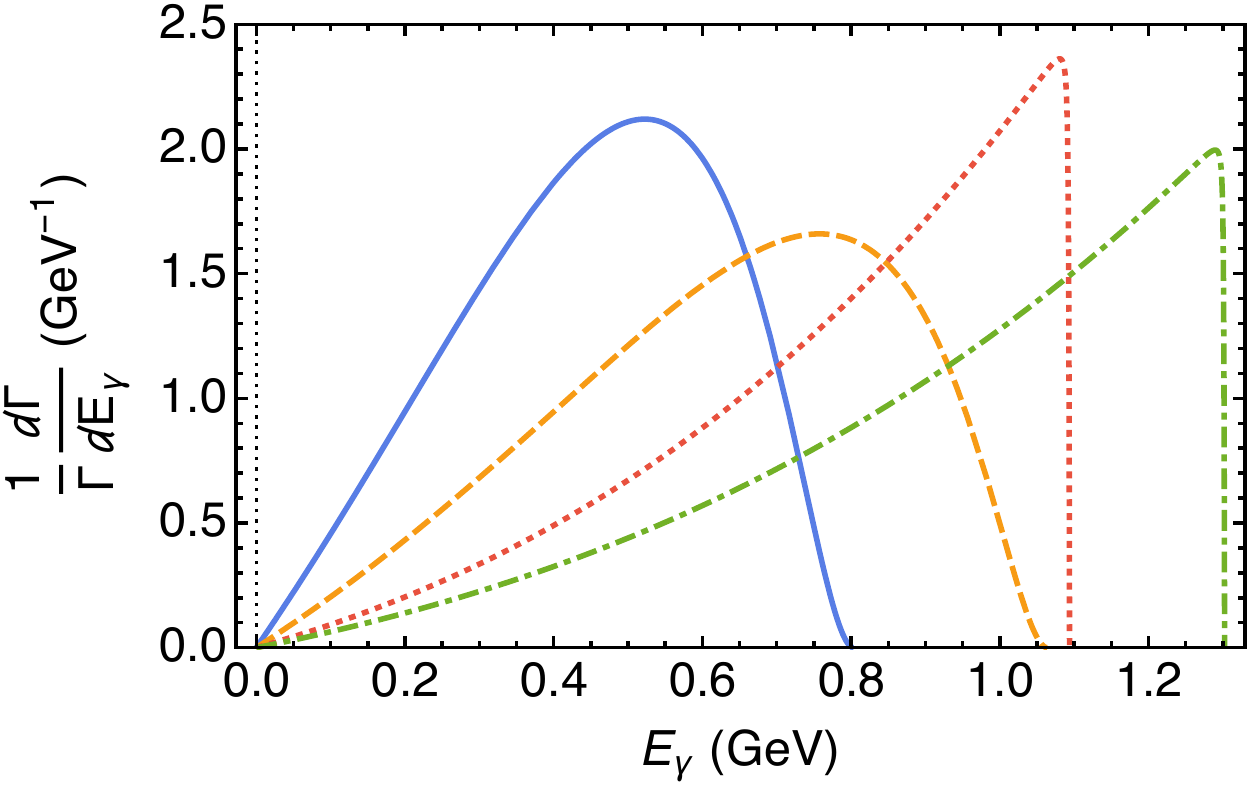} \label{diffdecayaxialcc}}
\captionsetup{singlelinecheck=off}
    \caption[]{Differential decay rates as functions of photon 
    energy $E_\gamma$ for axial operators (from \cite{Hazard:2016fnc}). Plotted decay rates are for
    (a) $\Upsilon(1S) \to\gamma \mu \tau$ or $ \gamma e \tau$ (solid blue), 
    $\Upsilon(2S) \to \gamma \mu \tau$ or $\gamma e \tau$ (short-dashed gold), 
    $\Upsilon(3S) \to \gamma \mu \tau$ or $\gamma e \tau$ (dotted red),
    $\Upsilon(1S) \to \gamma e \mu$ (dot-dashed green),
    $\Upsilon(2S) \to \gamma e \mu$ and $\Upsilon(3S) \to \gamma e \mu$ (long-dashed purple);
    (b) $J/\psi \to \gamma \mu \tau$ or $\gamma e \tau$ (solid blue),
    $\psi(2S) \to \gamma \mu \tau$ or $ \gamma e \tau$ (short-dashed gold), 
    $J/\psi \to \gamma e \mu$ (dotted red),
    $\psi(2S) \to \gamma e \mu$ (dot-dashed green).}
\label{differentialdecayaxial}
\end{figure}

\begin{figure}
\subfigure[]{\includegraphics[scale=.63]{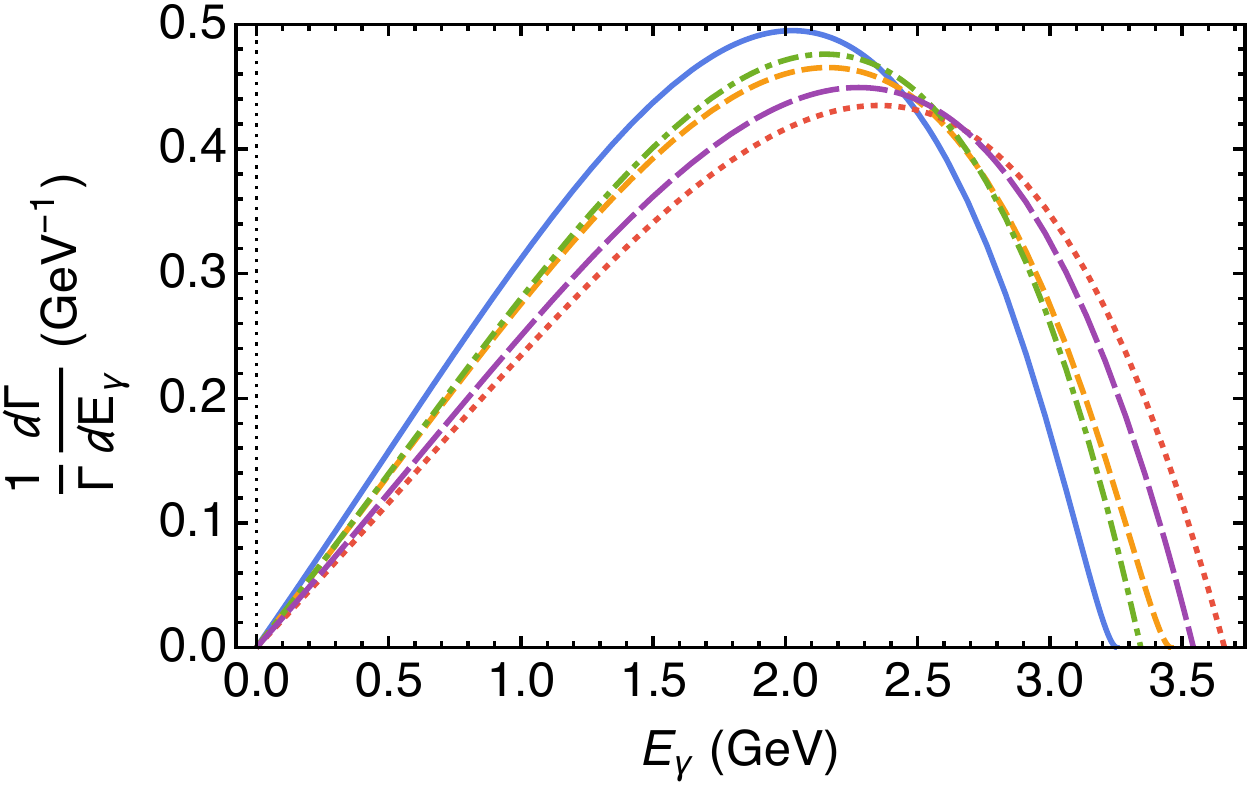} \label{diffdecayscalarbb}}
\subfigure[]{\includegraphics[scale=.63]{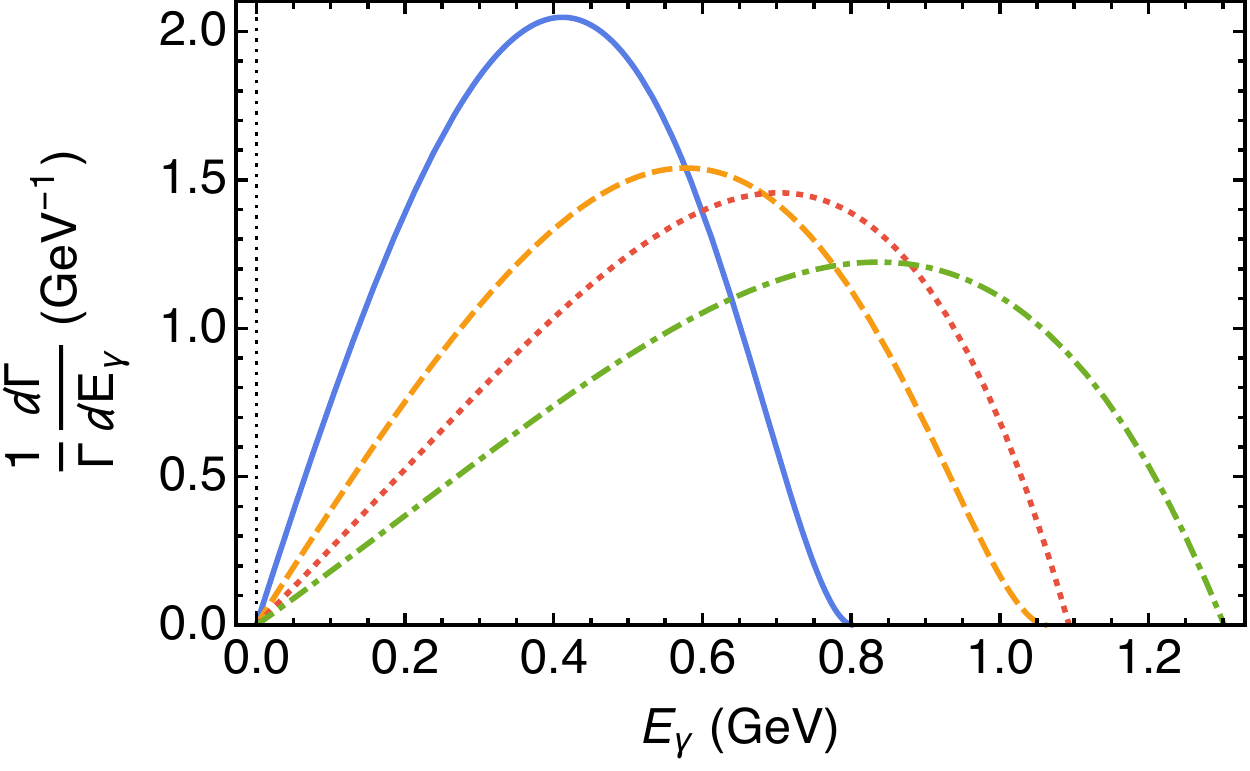} \label{diffdecayscalarcc}}
\captionsetup{singlelinecheck=off}
    \caption[]{Differential decay rates as functions of photon 
    energy $E_\gamma$ for scalar/pseudo-scalar operators (from \cite{Hazard:2016fnc}). Plotted decay rates are for
    (a) $\Upsilon(1S) \to \gamma \mu \tau$ or $\gamma e \tau$ (solid blue), 
    $\Upsilon(2S) \to \gamma \mu \tau$ or $ \gamma e \tau$ (short-dashed gold), 
    $\Upsilon(3S) \to \gamma \mu \tau$, $\gamma e \tau$, or $\gamma e \mu$ (dotted red),
    $\Upsilon(1S) \to \gamma e \mu$ (dot-dashed green),
    $\Upsilon(2S) \to \gamma e \mu$ (long-dashed purple);
    (b) $J/\psi \to \gamma \mu \tau$ or $\gamma e \tau$ (solid blue),
    $\psi(2S) \to \gamma \mu \tau$ or $ \gamma e \tau$ (short-dashed gold), 
    $J/\psi \to \gamma e \mu$ (dotted red),
    $\psi(2S) \to \gamma e \mu$ (dot-dashed green).}
\label{differentialdecayscalar}
\end{figure}

Here $m_{12}^2 = (p_1+p_2)^2$ \cite{PDG}, where $p_{1}$ and $p_{1}$ are the momenta of $\ell_1$ and $\ell_2$.  The WC indices are suppressed for brevity (i.e. $C_{SR(L)}^{q \ell_1 \ell_2} \to C_{SR(L)}$) and the remaining constants are defined in Section \ref{sec:2bodyVec}.  Integrating Eq. \ref{3bodydiffdecayrates} over $m_{12}^2$ gives the total decay rates

\begin{eqnarray} \label{3bodydecayrate}
\Gamma_A (V \to \gamma \ell_1 \overline \ell_2) &=& \frac{1}{18} \frac{\alpha Q_q^2}{\left(4 \pi\right)^2} 
\frac{f_V^2 m_V^3}{\Lambda^4} \left(C_{AL}^2+C_{AR}^2 \right) f(y^2), 
\nonumber \\
\Gamma_S(V \to \gamma \ell_1 \overline \ell_2) &=& \frac{1}{144} \frac{\alpha Q_q^2}{\left(4 \pi\right)^2} 
\frac{f_V^2 G_F^2 m_V^7}{\Lambda^4} \left(C_{SL}^2+C_{SR}^2 \right) y^2 f(y^2), 
\\
\Gamma_P (V \to \gamma \ell_1 \overline \ell_2) &=& \frac{1}{144} \frac{\alpha Q_q^2}{\left(4 \pi\right)^2} 
\frac{f_V^2 G_F^2 m_V^7}{\Lambda^4} \left(C_{PL}^2+C_{PR}^2 \right) y^2 f(y^2),
\nonumber
\end{eqnarray}

where $f(y^2) = 1-6y^2-12y^4\text{log}\left(y\right)+3y^4+2y^6$.  Normalizing the differential decay rates to their total rate cancels out the unknown WCs allowing us to plot the normalized differential decay distributions as a function of photon energy, $E_{\gamma}$, for the axial operators in Fig. \ref{differentialdecayaxial} and the scalar or pseudo-scalar operators in Fig. \ref{differentialdecayscalar}.

Currently there are no experimental constraints on radiative lepton flavor violating decays of vector quarkonia.  This means we cannot yet place constraints on the WCs involved in their transitions.

\section{Conclusion} 

Lepton flavor violating transitions provide a powerful engine for new physics searches.  Any new physics model with flavor violation at high scales can be cast in terms of the ${\cal L}_{\rm eff}$ at low energies. Two-body decays allow for operator selection and reduce our reliance on the single operator dominance assumption while radiative lepton flavor violating decays can provide complimentary access to the same operators.

\section{Acknowledgments} 

I am grateful to my advisor, Alexey A. Petrov, for all of his help in completing this project.

\end{document}